\documentclass[preprint, 11pt]{elsarticle}
\usepackage{hyperref}

\usepackage{amssymb}
\def\Eins{{\mathchoice {\rm 1\mskip-4mu l} {\rm 1\mskip-4mu l}
{\rm 1\mskip-4.5mu l} {\rm 1\mskip-5mu l}}}
\begin{document}

\begin{frontmatter}

\title{A covariant representation of the Ball-Chiu vertex\footnote{Preprint AEI-2012-104}}

\author[ifm,bologna]{Naser Ahmadiniaz}
\ead{naser@ifm.umich.mx}

\author[ifm,bologna,aei]{Christian Schubert}
\ead{schubert@ifm.umich.mx}

\address[ifm]{
Instituto de F\'{\i}sica y Matem\'aticas,
Universidad Michoacana de San Nicol\'as de Hidalgo,
Edificio C-3, Apdo. Postal 2-82,
C.P. 58040, Morelia, Michoac\'an, Mexico
}

\address[bologna]{
Dipartimento di Fisica, Universit\`a di Bologna and INFN, Sezione di Bologna,
Via Irnerio 46, I-40126 Bologna, Italy
}

\address[aei]{
Max-Planck-Institut f\"ur Gravitationsphysik, Albert-Einstein-Institut,
M\"uhlenberg 1, D-14476 Potsdam, Germany
}


\begin{abstract}
In nonabelian gauge theory the three-gluon vertex function contains important structural information, 
in particular on infrared divergences, and is also an essential ingredient in the Schwinger-Dyson equations. 
Much effort has gone into analyzing its general structure, and at the one-loop level also a number of 
explicit computations have been done, using various approaches. 
Here we use the string-inspired formalism to unify the calculations of the scalar, spinor and gluon 
loop contributions to the one-loop vertex, leading to an extremely compact representation in all cases. 
The vertex is computed fully off-shell and in dimensionally continued form, so that it can be used as a
building block for higher-loop calculations. We find that the Bern-Kosower loop replacement rules, 
originally derived for the on-shell case, hold off-shell as well. We explain the relation of the structure 
of this representation to the low-energy effective action, and establish the precise connection with 
the standard Ball-Chiu decomposition of the vertex. This allows us also to predict that the
vanishing of the completely antisymmetric coefficient function S of this decomposition
is not a one-loop accident, but persists at higher loop orders. The sum rule found by Binger and Brodsky, 
which leads to the vanishing of the one-loop vertex in $N=4$ SYM theory, in the present approach relates 
to worldline supersymmetry. 
\end{abstract}

 
\end{frontmatter}

\newcommand{\be}{\begin{equation}}
\newcommand{\ee}{\end{equation}\noindent}
\newcommand{\bear}{\begin{eqnarray}}
\newcommand{\ear}{\end{eqnarray}\noindent}
\newcommand{\no}{\noindent}
\newcommand{\non}{\nonumber\\}

\def\eq#1{{eq. (\ref{#1})}}
\def\eqs#1#2{{eqs. (\ref{#1}) -- (\ref{#2})}}

--------

\def\mn{{\mu\nu}}
\def\rs{{\rho\sigma}}
\newcommand{\Det}{{\rm Det}}
\def\Tr{{\rm Tr}\,}
\def\tr{{\rm tr}\,}
\def\sumij{\sum_{i<j}}
\def\e{\,{\rm e}}

\def\half{\frac{1}{2}}
\def\freeexp{{\rm e}^{-\int_0^Td\tau {1\over 4}\dot x^2}}
\def\kinb{{1\over 4}\dot x^2}
\def\kinf{{1\over 2}\psi\dot\psi}
\def\expk{{\rm exp}\biggl[\,\sum_{i<j=1}^4 G_{Bij}p_i\cdot p_j\biggr]}
\def\expp{{\rm exp}\biggl[\,\sum_{i<j=1}^4 G_{Bij}p_i\cdot p_j\biggr]}
\def\expshort{{\e}^{\half G_{Bij}p_i\cdot p_j}}
\def\expabb{{\e}^{(\cdot )}}
\def\epseps#1#2{\varepsilon_{#1}\cdot \varepsilon_{#2}}
\def\epsk#1#2{\varepsilon_{#1}\cdot p_{#2}}
\def\kk#1#2{p_{#1}\cdot p_{#2}}
\def\G#1#2{G_{B#1#2}}
\def\Gp#1#2{{\dot G_{B#1#2}}}
\def\GF#1#2{G_{F#1#2}}
\def\Dab{{(x_a-x_b)}}
\def\Dsq{{({(x_a-x_b)}^2)}}
\def\PITD{{(4\pi T)}^{-{D\over 2}}}
\def\4piTD{{(4\pi T)}^{-{D\over 2}}}
\def\4piT4{{(4\pi T)}^{-2}}
\def\TintmD{{\dps\int_{0}^{\infty}}{dT\over T}\,e^{-m^2T}
    {(4\pi T)}^{-{D\over 2}}}
\def\Tintm4{{\dps\int_{0}^{\infty}}{dT\over T}\,e^{-m^2T}
    {(4\pi T)}^{-2}}
\def\Tintm{{\dps\int_{0}^{\infty}}{dT\over T}\,e^{-m^2T}}
\def\Tint{{\dps\int_{0}^{\infty}}{dT\over T}}
\def\np{n_{+}}
\def\nm{n_{-}}
\def\Np{N_{+}}
\def\Nm{N_{-}}
\newcommand{\slG}{{{\dot G}\!\!\!\! \raise.15ex\hbox {/}}}
\newcommand{\Gd}{{\dot G}}
\newcommand{\Gund}{{\underline{\dot G}}}
\newcommand{\Gdd}{{\ddot G}}
\def\GBd12{{\dot G}_{B12}}
\def\Dx{\dps\int{\cal D}x}
\def\Dy{\dps\int{\cal D}y}
\def\Dpsi{\dps\int{\cal D}\psi}
\def\dint#1{\int\!\!\!\!\!\int\limits_{\!\!#1}}
\def\ddtau{{d\over d\tau}}
\def\ie{\hbox{$\textstyle{\int_1}$}}
\def\iz{\hbox{$\textstyle{\int_2}$}}
\def\id{\hbox{$\textstyle{\int_3}$}}
\def\ldop{\hbox{$\lbrace\mskip -4.5mu\mid$}}
\def\rdop{\hbox{$\mid\mskip -4.3mu\rbrace$}}
%
\newcommand{\1}{{\'\i}}
\def\dps{\displaystyle}
\def\sy{\scriptscriptstyle}
\def\sy{\scriptscriptstyle}

\def\del{\partial}
\def\deli{\partial_{\kappa}}
\def\delj{\partial_{\lambda}}
\def\delk{\partial_{\mu}}
\def\delij{\partial_{\kappa\lambda}}
\def\delik{\partial_{\kappa\mu}}
\def\deljk{\partial_{\lambda\mu}}
\def\delki{\partial_{\mu\kappa}}
\def\delkl{\partial_{\mu\nu}}
\def\delijk{\partial_{\kappa\lambda\mu}}
\def\deljkl{\partial_{\lambda\mu\nu}}
\def\delikl{\partial_{\kappa\mu\nu}}
\def\delijkl{\partial_{\kappa\lambda\mu\nu}}
\def\delijklm{\partial_{\kappa\lambda\mu\nu o}}
\def\O(#1){O($T^#1$)} 
\def\O2{O($T^2$)}
\def\O3{O($T^3$)}
\def\O4{O($T^4)}
\def\O5{O($T^5$)}
\def\dA{\partial^2}
\def\DA{\sqsubset\!\!\!\!\sqsupset}
\def\eins{  1\!{\rm l}  }

\newcommand{\Vka}{V_{\kappa}}
\newcommand{\Vla}{V_{\lambda}}
\newcommand{\Vmu}{V_{\mu}}
\newcommand{\Vnu}{V_{\nu}}
\newcommand{\Vro}{V_{\rho}}
\newcommand{\Vkala}{V_{\kappa\lambda}}
\newcommand{\Vkamu}{V_{\kappa\mu}}
\newcommand{\Vkanu}{V_{\kappa\nu}}
\newcommand{\Vlamu}{V_{\lambda\mu}}
\newcommand{\Vlanu}{V_{\lambda\nu}}
\newcommand{\Vlaka}{V_{\lambda\kappa}}
\newcommand{\Vmunu}{V_{\mu\nu}}
\newcommand{\Vmuka}{V_{\mu\kappa}}
\newcommand{\Vnuro}{V_{\nu\rho}}
\newcommand{\Vkalamu}{V_{\kappa\lambda\mu}}
\newcommand{\Vkalanu}{V_{\kappa\lambda\nu}}
\newcommand{\Vkalaro}{V_{\kappa\lambda\rho}}
\newcommand{\Vkamunu}{V_{\kappa\mu\nu}}
\newcommand{\Vlamunu}{V_{\lambda\mu\nu}}
\newcommand{\Vmunuro}{V_{\mu\nu\rho}}
\newcommand{\Vkalamunu}{V_{\kappa\lambda\mu\nu}}
\newcommand{\Fkala}{F_{\kappa\lambda}}
\newcommand{\Fkanu}{F_{\kappa\nu}}
\newcommand{\Flaka}{F_{\lambda\kappa}}
\newcommand{\Flamu}{F_{\lambda\mu}}
\newcommand{\Fmunu}{F_{\mu\nu}}
\newcommand{\Fnumu}{F_{\nu\mu}}
\newcommand{\Fnuka}{F_{\nu\kappa}}
\newcommand{\Fmuka}{F_{\mu\kappa}}
\newcommand{\Fkalamu}{F_{\kappa\lambda\mu}}
\newcommand{\Flamunu}{F_{\lambda\mu\nu}}
\newcommand{\Flanumu}{F_{\lambda\nu\mu}}
\newcommand{\Fkamula}{F_{\kappa\mu\lambda}}
\newcommand{\Fkanumu}{F_{\kappa\nu\mu}}
\newcommand{\Fmulaka}{F_{\mu\lambda\kappa}}
\newcommand{\Fmulanu}{F_{\mu\lambda\nu}}
\newcommand{\Fmunuka}{F_{\mu\nu\kappa}}
\newcommand{\Fkalamunu}{F_{\kappa\lambda\mu\nu}}
\newcommand{\Flakanumu}{F_{\lambda\kappa\nu\mu}}

\section{Introduction}
\label{intro}
\renewcommand{\theequation}{1.\arabic{equation}}
\setcounter{equation}{0}

The one-particle-irreducible (`1PI') off-shell three-gluon Green's function (in the following simply called ``three-vertex'' or
``vertex'')
is a basic object of interest in nonabelian gauge theory and quantum chromodynamics. 
It contains important structural information, in particular on infrared divergences (see, e.g., \cite{alhusc}) and references
therein) and is a main ingredient of the Schwinger-Dyson equations. 
In perturbation theory it can, computed explicitly to a certain loop order,  
in principle be used as a convenient building block for higher-loop calculations. 

\begin{figure}[h]
\hspace{100pt}{\centering
\includegraphics{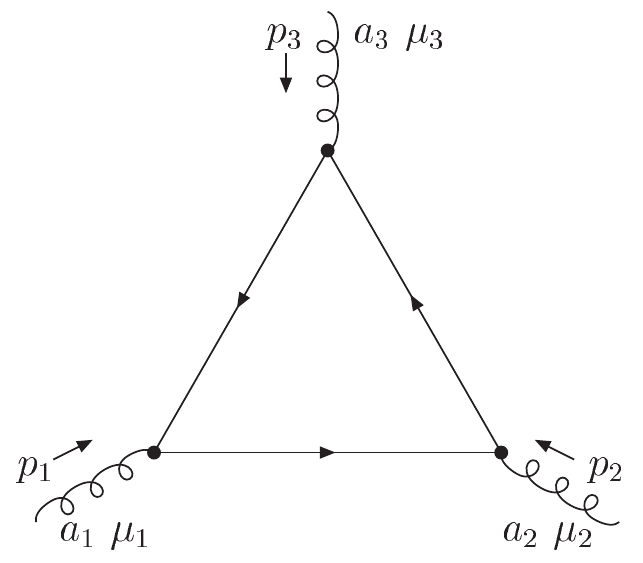}
}
\caption{Three-gluon vertex.}
\label{fig1}
\end{figure}

However, explicit calculations of the three-vertex have so far been essentially restricted to the one-loop
level \cite{celgon,pastar,balchi2,daosta-3gluonD,daossa-3gluonDm,binbro} (at two loops, the three-gluon vertex has been
obtained so far only for some very special momentum configurations \cite{daosta-2loop,davosl,gracey}). 
In this paper we will recalculate, in a simple and unifying way, the scalar, spinor and gluon loop
contributions to the one-loop three-vertex (with ``gluon'' we mean any nonabelian gauge boson).
In fig. \ref{fig1} for definiteness we show the fermion loop contribution. 

\noindent
Following the notation of \cite{daosta-3gluonD,daossa-3gluonDm}, we write

\bear
\Gamma_{\mu_1\mu_2\mu_3}^{a_1a_2a_3}(p_1,p_2,p_3) = -igf^{a_1a_2a_3}\Gamma_{\mu_1\mu_2\mu_3}(p_1,p_2,p_3)\, .
\label{stripcol}
\ear
The gluon momenta are ingoing and $p_1+p_2+p_3=0$. 
There are actually two diagrams differing by the two inequivalent orderings of the three gluons along the loops
(or equivalently by a change of the fermion line orientation). Those diagrams add to produce a factor of two.

By an analysis of the nonabelian gauge Ward identities, Ball and Chiu \cite{balchi2} in 1980 found a
form factor decomposition of this vertex which is valid at any order in perturbation theory, 
and also independent of whether the particle in the loop is a scalar, fermion or gluon, with the only restriction that
a covariant gauge be used. It can be written as

\bear
\Gamma_{\mu_{1}\mu_{2}\mu_{3}}(p_{1},p_{2},p_{3})&=&
A(p_1^2,p_2 ^2;p_3^2)g_{\mu_{1}\mu_{2}}(p_1-p_2)_{\mu_{3}}
+ B(p_1^2,p_2 ^2;p_3^2)g_{\mu_{1}\mu_{2}}
(p_1+p_2)_{\mu_{3}}\non
&&+ C(p_1^2,p_2 ^2;p_3^2)\Big\lbrack p{_1}_{\mu_2}p{_2}_{\mu_1} - p_1\cdot p_2\, g_{\mu_{1}\mu_{2}} \Bigl\rbrack(p_1-p_2)_{\mu_3}\non
&&+\frac{1}{3}S( p_1^2,p_2 ^2,p_3^2)\Bigl\lbrack p{_1}_{\mu_3}p{_2}_{\mu_1}p{_3}_{\mu_2}+p{_1}_{\mu_2}p{_2}_{\mu_3}p{_3}_{\mu_1}\Bigl\rbrack\non
&&+F( p_1^2,p_2 ^2;p_3^2)\Bigl\lbrack {p_1}_{\mu_2}p{_2}_{\mu_1}\ - p_1\cdot p_2 \, g_{\mu_1\mu_2} \Bigr\rbrack
\Bigl\lbrack
p_1\cdot p_3 \, p{_2}_{\mu_3}
-
p_2\cdot p_3\, p{_1}_{\mu_3}
\Bigl\rbrack
\non
&& + H(p_1^2,p_2 ^2,p_3^2)
\Big\{-g_{\mu_1\mu_2}\Bigl\lbrack p{_1}_{\mu_3} p_2\cdot p_3 - p{_2}_{\mu_3} p_1\cdot p_3\Bigl\rbrack\non
&&\qquad\qquad\qquad\quad +\frac{1}{3}\Big\lbrack p{_1}_{\mu_3}p{_2}_{\mu_1}p{_3}_{\mu_2}-p{_1}_{\mu_2}p{_2}_{\mu_3}p{_3}_{\mu_1}\Bigl\rbrack\Big\}\non
&&+\Big\{ {\mbox {cyclic permutations of } (p_1,\mu_1),(p_2,\mu_2),(p_3,\mu_3)}\Big\} \, .
\non
\label{B.C}
\ear
Here the functions $A$, $C$ and $F$ are symmetric in the first two arguments, 
the function $B$ is antisymmetric in the first two arguments, 
$H$ is totally symmetric and $S$ totally antisymmetric with respect to interchange of any pair of arguments. 
A different decomposition of the three-gluon vertex was proposed in \cite{kimbak}.

Although we wish to study the off-shell vertex, with our calculation method
it will be convenient to contract it with polarization vectors $\varepsilon_{1,2,3}$.
Those vectors are arbitrary and serve book-keeping purposes only. Thus we will use
(\ref{B.C}) in the form

\bear
\varepsilon_1^{\mu_1}\varepsilon_2^{\mu_2}\varepsilon_3^{\mu_3}
\Gamma_{\mu_{1}\mu_{2}\mu_{3}}(p_{1},p_{2},p_{3})&\equiv & 
\Bigl\lbrace A(p_{1}^2,p_2^2;p_3^2)T_{A}+B(p_{1}^2,p_2^2;p_3^2)T_{B}
+C(p_{1}^2,p_2^2;p_3^2)T_{C} \non&&
+ F(p_{1}^2,p_2^2;p_3^2)T_{F} \Bigr\rbrace +
\left \lbrace {\mbox {2 cyclic permutations}} \right\rbrace
\non&&
+H(p_{1}^2,p_2^2,p_3^2)T_{H}+S(p_{1}^2,p_2^2,p_3^2)T_{S} \, .
\non
\label{vertexcontracted}
\ear
Note that the tensors $T_F$ and $T_H$ 
are totally transversal, i.e., they give zero when any $\varepsilon_i$ is replaced by $p_i$. 

For the gluon loop in Feynman gauge and at the one-loop level, Ball and Chiu also calculated the coefficient 
functions $A$ to $H$, up to their constant terms in the $\epsilon$-expansion; here it turned out that $S$ actually vanishes.
Later Davydychev, Osland and Tarasov \cite{daosta-3gluonD} computed this gluon loop contribution  
vertex more generally for an arbitrary covariant gauge, and in arbitrary spacetime dimension.

The quark loop contribution to the vertex was first calculated for massless quarks and in the symmetric limit $p_1^2=p_2^2=p_3^2$
by Celmaster and Gonsalves \cite{celgon} and Pascual and Tarrach \cite{pastar}. For general off-shell momenta the
massless quark loop contribution was obtained in \cite{daosta-3gluonD}, and the massive quark loop one in \cite{daossa-3gluonDm}
(again in arbitrary spacetime dimension). 

However, this is not yet the whole story, since for the gluon loop contribution to the vertex there are subtle issues
with gauge dependence. When calculated in the standard formalism using any covariant gauge, it satisfies rather complicated
Slavnov-Taylor identities involving not only the gluon propagator, but also the ghost propagator and the gluon-ghost-ghost vertex
(see, e.g., \cite{marpag,balchi2,binbro}). The scalar and fermion loop contributions, on the other hand, 
satisfy the simple QED-like Ward
identity 

\bear
\Gamma (\varepsilon_3 \to p_3) &=& 
-(p_1^2\varepsilon_1\cdot\varepsilon_2-p_1\cdot\varepsilon_1 p_1\cdot\varepsilon_2)
\Bigl(1-\Pi(p_1^2)\Bigr)
\nonumber\\
&& +(p_2^2\varepsilon_1\cdot\varepsilon_2-p_2\cdot\varepsilon_1 p_2\cdot\varepsilon_2)
\Bigl(1-\Pi(p_2^2)\Bigr) 
\nonumber\\
\label{ward}
\ear
where $\Pi (p^2)$ is the corresponding vacuum polarization function.
Having the same simple Ward identity also for the gluon loop case is possible, but requires 
more sophisticated techniques. It can be achieved using either the background field method (`BFM')
\cite{abbott,abgrsc,ditreubook-selected_topics} with Feynman gauge for the quantum field, or the pinch technique \cite{corpap,binpap-rev}. 
Although very different, those two methods turn out to lead to precisely the same Green's functions \cite{dewedi,papavassiliou95}.
The corresponding three-gluon vertex, also called the ``gauge-invariant vertex'', was studied by Freedman et al. \cite{fgjr}
with an emphasis on its conformal properties. 
Binger and Brodsky \cite{binbro} explicitly calculated it in the 
transversality-based Ball-Chiu decomposition (\ref{B.C}), as well as in
a different basis related to current conservation. 
They also calculated the scalar loop contribution, which enabled them to find certain 
sum rules between the massless scalar, spinor and gluon loop contributions.
In particular, with all particles massless and in the adjoint representation they obtain the identity

\bear
3 \Gamma_0 + 2 \Gamma_{\half} + \Gamma_1   = 0
\label{relsusy}
\ear
(here we introduce the convention that a subscript $s=0$ refers to a scalar loop,
$s=\half$ to a spinor loop, and $s=1$ to a gluon loop including the ghost loop
contribution).
This identity is related to supersymmetry, and implies that the off-shell three-gluon
amplitude in $N=4$ Super-Yang-Mills theory vanishes. It generalizes the well-known vanishing of the 
gluon self-energy in that theory, but contrary to that fact does not obviously relate to the finiteness of
the theory.
 
In the present work, we will recalculate the scalar, spinor and gluon loop contributions to this ``gauge-invariant''
three-gluon vertex using the ``string-inspired'' formalism along the lines of 
\cite{berkos:prl,berkos:npb362,berkos:npb379,strassler,strassler2,5,26} 
(for a review, see \cite{41}). 
Our starting point is the ``Bern-Kosower master formula'' \cite{berkos:prl,berkos:npb362,berkos:npb379}:
 
\bear
&& \Gamma^{a_{1}\dots a_{N}}_0[p_{1},\varepsilon_{1};\dots;p_{N},\varepsilon_{N}]\non
 &&=(-ig)^{N}\mbox{tr}(T^{a_{1}}\dots T^{a_{N}})(2\pi)^{D}i\delta(\sum p_{i})\int_{0}^{\infty} dT(4\pi T)^{-D/2}e^{-m^2 T}\non
 && \times\int_{0}^{T}d\tau_{1}\int_0^{\tau_{1}}d\tau_2\dots\int_0^{\tau_{N-2}}d\tau_{N-1}
 \nonumber\\ && \times
 \exp\Bigg\{\sum_{i,j=1}^N\left[\frac{1}{2}
  G_{Bij}p_{i}\cdot p_{j}
-i\dot{G}_{Bij}\varepsilon_{i}\cdot p_{j}+\frac{1}{2}\ddot{G}_{Bij}\varepsilon_{i}\cdot\varepsilon_{j}\right]\Bigg\}
\Biggl\vert_{\rm lin (\varepsilon_1 \ldots \varepsilon_N)} \, .
\nonumber\\
\label{bk}
\ear
As it stands, this formula represents the color-ordered contribution to the 
1PI $N$ - gluon amplitude due to a (complex) scalar loop
of mass $m$, calculated in $D$ spacetime dimensions. The $i$th gluon carries the momentum $p_i$, polarization $\varepsilon_i$ 
and a gauge group generator $T^{a_i}$ in some representation. 
$T$ is the total proper-time length of the loop, and $\tau_i$ is the position in proper-time
along the loop of gluon $i$.  One integration is redundant and has been eliminated by setting $\tau_N=0$. 
The derivation of this formula involved a formal exponentiation, which needs to be undone by expanding out the exponential
factor and keeping only the terms linear in each of the $N$ polarization vectors.
The color-ordering means that one still has to sum over all $(N-1)!$ inequivalent orderings of the gluons along the loop to get the
full amplitude.  
$G_{Bij}\equiv G_B(\tau_i,\tau_j)$ denotes the ``bosonic'' worldline Green's function, defined by 

\be
G_B(\tau_1,\tau_2)=\mid \tau_1-\tau_2\mid 
-{{(\tau_1-\tau_2)}^2\over T}, 
\label{defG}
\ee
\no
and dots generally denote a
derivative acting on the first variable.
Explicitly,

\begin{eqnarray}
\dot G_B(\tau_1,\tau_2) &=& {\rm sign}(\tau_1 - \tau_2)
- 2 {{(\tau_1 - \tau_2)}\over T}\, ,\nonumber\\
\ddot G_B(\tau_1,\tau_2)
&=& 2 {\delta}(\tau_1 - \tau_2)
- {2\over T}\, .\quad \nonumber\\
\label{GdGdd}
\end{eqnarray}
\noindent

The master formula (\ref{bk}) was originally derived 
from string theory  \cite{berkos:prl,berkos:npb362,berkos:npb379}, starting from a representation of the $N$-gluon  
amplitude for the heterotic string and analyzing its field theory limit. Here the object of interest was, however, the full $N$-gluon 
amplitude on-shell, rather than the 1PI amplitude off-shell. Thus on one hand on-shell conditions were used from the beginning, already at the
string level; 
on the other hand the fact that the distinction between reducible and irreducible diagrams emerges only in the
field theory limit made it possible to establish certain formal rules that allow one to reconstruct, from the formula for the 1PI amplitude
(\ref{bk}), also all the missing reducible contributions to the full on-shell matrix element. 
Bern and Kosower were moreover able to derive simple ``loop replacement rules'', based on worldsheet supersymmetry, that allow one
to obtain from (\ref{bk}) also integral representations for the spinor and gluon loop contributions to the full on-shell
$N$-gluon amplitudes \cite{berkos:prl,berkos:npb362,berkos:npb379}. 
We need not discuss these ``Bern-Kosower rules'' here in full, 
but it is important to note that they all involve integration-by-parts (`IBP') in an essential way. 
Namely, performing the expansion of the exponential factor in (\ref{bk}) will yield an integrand
$\sim P_N\,e^{(\cdot)}$, where we abbreviated

\bear 
e^{(\cdot)}:= \mbox{exp}\Bigg\{\frac{1}{2}\sum_{i,j=1}^{N} G_{Bij}p_{i}\cdot p_{j}\Bigg\} \, ,
\label{abbrev}
\ear
and $P_N$ is a polynomial in $\dot G_{Bij},\ddot G_{Bij}$ and the kinematic invariants. 
It is possible to remove 
all second derivatives $\ddot{G}_{Bij}$ appearing in $P_N$ by suitable integrations-by-parts,
leading to a new integrand $\sim Q_N\,e^{(\cdot)}$ which is the real starting point for the
application of the Bern-Kosower rules. Relevant for the following will be only the
``loop replacement rules''. The rule for passing from the scalar to the spinor loop is easy
to state in general: look in $Q_N$ for ``$\tau$-cycles'', that is, 
products of $\dot G_{Bij}$'s whose indices form a closed chain. A $\tau$-cycle can
thus be written as 
$\dot G_{Bi_1i_2} \dot G_{Bi_2i_3} \cdots \dot G_{Bi_ni_1}$
(to put it into this form may require the use of the antisymmetry of $\dot G_B$, e.g.
$\dot G_{B12}\dot G_{B12}=-\dot G_{B12}\dot G_{B21}$).
Then, apart from a global factor of $-2$ correcting
for degrees of freedom and statistics, the integrand for the spinor
loop case can be obtained from the one for
the scalar loop simply
by simultaneously replacing every $\tau$-cycle appearing in $Q_N$ by 

\begin{equation}
\dot G_{Bi_1i_2} 
\dot G_{Bi_2i_3} 
\cdots
\dot G_{Bi_ni_1}
\rightarrow 
\dot G_{Bi_1i_2} 
\dot G_{Bi_2i_3} 
\cdots
\dot G_{Bi_ni_1}
-
G_{Fi_1i_2}
G_{Fi_2i_3}
\cdots
G_{Fi_ni_1}\, ,
\nonumber\\
\label{subrule}
\end{equation}
\no
where $G_{F12}\equiv {\rm sign}(\tau_1-\tau_2)$ denotes the `fermionic' worldline Green's function. 
The rule for passing from the scalar to the gluon loop is similar but somewhat more complicated, and will
be dealt with in section \ref{section-gluon} below. 

This formalism was used for the first calculation of the one-loop on-shell QCD five-gluon amplitudes 
\cite{bediko5glu}, but not further employed for such on-shell multi-gluon amplitude calculations 
due to the emergence of other extremely powerful methods for the computation of one-loop on-shell 
amplitudes such as generalized unitarity; see, e.g., \cite{bediko:annphys}.

In the present paper, we will instead start an effort to exploit the Bern-Kosower formalism 
as a tool for the calculation of the one-loop $N$-gluon vertex.  
In the original string-based formalism going off-shell is highly nontrivial, although not impossible.
In \cite{dlmmr:plb351,dlmmr:npb469} a formalism was developed that, in principle, allows one to obtain the one-loop off-shell
gluon amplitudes in pure Yang-Mills theory from the open string, and it was shown to correctly reproduce the renormalization
constants for the two-, three- and four-point vertices. Here, however, we will use the simpler
approach to the Bern-Kosower formalism due to Strassler \cite{strassler,strassler2}, which uses
string theory only as a guiding principle. 
The starting point of this ``string-inspired worldline formalism'' is the following path integral representation of the
nonabelian one-loop effective action due to a scalar loop \cite{strassler} (this generalizes
Feynman's famous 1950 formula for scalar QED \cite{feynman:pr80}),

 \begin{equation}
\Gamma_{0}\lbrack A\rbrack   =  {\dps\int_0^{\infty}}
{dT\over T}\,
e^{-m^2T}
{\rm tr}{\dps\int} {\cal D} x
\exp\Bigl [- \int_0^T d\tau
\Bigl (\frac{1}{4}{\dot x}^2 
+ ig\dot x\cdot A
\Bigr )\Bigr ] \, ,
\label{avpi}
\end{equation}
\no
where the integral ${\dps\int} {\cal D} x$ is over the space of all closed trajectories in spacetime with
periodicity $T$ in proper-time, $x^{\mu}(T) = x^{\mu}(0)$. 

Although this approach fell somewhat short of yielding a rederivation of the full set of Bern-Kosower
rules including the non-1PI contributions, it provides a simple way to see that the master formula
(\ref{bk}) as it stands is valid off-shell as a formula for the 1PI part of the $N$-gluon Green's function
(see \cite{strassler,41}).
In \cite{strassler2}, Strassler moreover started a systematic investigation of the IBP procedure, and
discovered that it bears an interesting connection to gauge invariance.
Namely, it so turns out that, once all 
$\ddot G_{Bij}$'s have been removed and all terms contributing
to a given `$\tau$ -- cycle' 
$\dot G_{Bi_1i_2} 
\dot G_{Bi_2i_3} 
\cdots
\dot G_{Bi_ni_1}$
been combined, the sum of their Lorentz factors can be written
as a `Lorentz cycle' $Z_n$, defined by

\bear
Z_2(ij)&\equiv&
\half {\rm tr}\Bigl(f_if_j\Bigr) \, ,
\non
Z_n(i_1i_2\ldots i_n)&\equiv&
{\rm tr}
\Bigl(
\prod_{j=1}^n
f_{i_j}\Bigr) 
\quad (n\geq 3)
\, ,\non
\label{defZn}
\ear\no
where

\bear
f_i^{\mu\nu}&\equiv&
p_i^{\mu}\varepsilon_i^{\nu}
- \varepsilon_i^{\mu}p_i^{\nu}
\label{deff}
\ear
is the momentum space form of the abelian field strength tensor.
$Z_n$ generalizes the transversal projector
which is familiar from the two-point case. 
However, in \cite{strassler2}
no systematic way was found to perform the partial
integrations at arbitrary $N$, nor how to preserve the permutation
symmetry. 
This issue was taken up again in \cite{26}, where a definite
and computerizable IBP algorithm was given which works for any $N$
and preserves the full permutation symmetry. 
This algorithm is still not satisfactory from the point of view of gauge
invariance, though. A given term in the integrand after the IBP in general
has not only cycle factors, but also a leftover, called ``tail'', and 
the algorithm arranges into field strength tensors only the polarization
vectors contained in the cycles, not the ones in the tails.
Only very recently an extension of the algorithm of \cite{26} was found
which, for any $N$ and preserving the permutation invariance, achieves this 
``covariantization'' for all the polarization vectors, including the ones in 
tails \cite{91}. 

This in some sense completes the investigation started in \cite{strassler2}. 
It also suggests that, with this optimized IBP at hand, the string-inspired formalism
might become a powerful tool for the computation of the $N$-vertex. This is for three reasons:
(i) The covariantization means that the bulk integrand after the IBP is manifestly transversal,
so that any nontransversality must come from boundary terms. Thus the IBP procedure itself
should generate a transversality-based form factor decomposition similar to the Ball-Chiu one (\ref{B.C}).
(ii) Like the Ball-Chiu one, this decomposition will respect the cyclic invariance (which is
the remnant of the permutation invariance after the color ordering). 
(iii) The work of \cite{strassler} also suggests that the ``loop-replacement'' part of the Bern-Kosower
rules may hold off-shell for the 1PI amplitudes, which would reduce the calculational effort very
significantly. 

Here we will recalculate the three-gluon vertex along the above lines, and find these
expectations to be fully justified. The organization of the paper is as follows: 

In section \ref{scalar} we will start with the scalar loop contribution to the vertex. We perform
the IBP using the old algorithm of \cite{26} as well as the improved one of \cite{91}. With both choices
we obtain a very compact integral representation for the vertex, however the new algorithm has the advantage 
that all nontransversality is pushed into the boundary (two-point) terms. 

In sections \ref{section-spinor} and \ref{section-gluon} 
we show that the ``loop replacement rules'' indeed hold for the three-gluon vertex. 
As is well-known, the Dirac fermion possesses an $N=1$ supersymmetric worldline path integral
representation \cite{brdiho,bermar,bssw,rajeev,polyakovbook,andtse,bacalu}, 
and the gluon an $N=2$ supersymmetric one \cite{brdiho,bssw,hppt,strassler,dahusi}. 
Analogously to the
original string-based derivation of those rules \cite{berkos:prl,berkos:npb362,berkos:npb379}, where 
worldsheet SUSY was identified as the underlying symmetry, in the worldline approach 
the same rules can be related to this worldline SUSY \cite{strassler,41}.

In section \ref{summary} we summarize and unify our results for the scalar, spinor and gluon loop.
In section \ref{comp} we establish their exact relation to the Ball-Chiu decomposition (\ref{B.C}),
and also explicitly verify the Ward identities (\ref{ward}) and the Binger-Brodsky relation (\ref{relsusy}).

The 1PI vertices hold the same information as the effective action. Nevertheless, contrary to the QED case where
there is no essential difference between the calculation of the effective action and of the
off-shell $N$-photon amplitudes, in the nonabelian case the effective action is mathematically an intrinsically 
more natural object. This is because it can be written in terms of full field strength tensors 

\bear
F_{\mn} \equiv F_{\mn}^a T^a = (\partial_{\mu}A_{\nu}^a - \partial_{\nu}A_{\mu}^a) T^a + ig[A_{\mu}^bT^b,A_{\nu}^cT^c]
\label{defF}
\ear
whereas upon Fourier transformation those inevitably get split up into their ``abelian parts''
$f_{\mn}^a:=\partial_{\mu}A_{\nu}^a - \partial_{\nu}A_{\mu}^a $ and the commutator terms.
This suggests that the analysis of the structure of the 1PI vertices should benefit from a comparison with the
low-energy expansion of the effective action, and indeed we will show in section \ref{ea} for the three-point case that 
in the present formalism, due to the systematic generation of ``abelian'' field strength tensors and   
commutator terms by the IBP, it is possible to keep the relation between the effective action and the 
vertex very transparent. 

Our conclusions are given in section \ref{conc}. In particular, 
we give there a general argument showing that the off-shell validity of the loop replacement rules 
extends to the $N$-vertex. 
\ref{conv} lists our conventions.

\section{The scalar loop case}
\label{scalar}
\renewcommand{\theequation}{2.\arabic{equation}}
\setcounter{equation}{0}

Before coming to the calculation of the (off-shell, 1PI) three-gluon amplitude for a scalar loop, let
us first consider the two-point (vacuum polarization) case. This will not only be useful as a warm-up,
but also for the verification of the Ward identitiy (\ref{ward}) later on. 

For $N=2$ we get from the master formula (\ref{bk}), after expanding out the exponential (in the following we generally
omit the global factor $(2\pi)^{4}i\delta(\sum p_i)$ for energy-momentum conservation),

\bear
\Gamma_{0}^{a_{1}a_{2}}[p_{1},\varepsilon_{1};p_{2},\varepsilon_2]
 &=&(-ig)^{2}\mbox{tr}(T^{a_{1}} T^{a_{2}})\int_{0}^{\infty} dT(4\pi T)^{-D/2}e^{-m^2 T}
 \non && \times
 \int_{0}^{T}d\tau_{1} \, (-i)^2 P_2 \,e^{G_{B12}p_1\cdot p_2}\non
\label{expand}
\ear
where

\bear
P_2 = 
\dot{G}_{B12}\varepsilon_{1}\cdot p_{2}\dot{G}_{B21}\varepsilon_{2}\cdot p_1
-\ddot{G}_{B12} \varepsilon_{1}\cdot\varepsilon_{2} \, .
\label{P2}
\ear
By an IBP of the term involving $\ddot G_{B12}$ we can remove the second derivative and
transform $P_2$ into $Q_2$, 

\bear
Q_2 &=&
\dot{G}_{B12}\dot{G}_{B21}
\bigl(\varepsilon_{1}\cdot p_{2}\varepsilon_{2}\cdot p_1- \varepsilon_1\cdot\varepsilon_2 p_1\cdot p_2 \bigr)
\non
&=& \frac{1}{2}\dot{G}_{B12}\dot{G}_{B21}\tr (f_1f_2)
= \dot{G}_{B12}\dot{G}_{B21}Z_2(12)
 \, .
\label{Q2}
\ear
Thus the IBP has allowed us to absorb the polarization vectors into the ``abelian'' field strength tensors
$f_i$, defined in (\ref{deff}), thereby making the transversality of the two-point function manifest.
Setting $p\equiv p_1=-p_2$ and using $\mbox{tr}(T^{a_{1}} T^{a_{2}})=C ( r)\delta^{a_1a_2}$, we can write

\bear
\Gamma_{0}^{a_{1}a_{2}}[p_{1},\varepsilon_{1};p_{2},\varepsilon_2]
&=& \varepsilon_{1}^{\mu}\Pi_{0\mn}^{a_1a_2}( p )\varepsilon_{2}^{\nu}\, , \non
\Pi_{0\mn}^{a_1a_2}( p ) &=& \delta^{a_1a_2}(\eta_{\mn}p^2-p_{\mu}p_{\nu})\Pi_0(p^2) 
\label{GammaPi}
\ear
where

\bear
\Pi_0(p^2) &=&  C(r)\frac{g^2}{(4\pi)^{D/2}}
\int_{0}^{\infty} \frac{dT}{T^{\frac{D}{2}}}e^{-m^2 T}\int_{0}^{T}d\tau_{1}
\dot{G}_{B12}\dot{G}_{B21} \,\e^{-G_{B12}p^2} 
\, .
\non
\label{Piscal}
\ear
Moving on to the three-point level, here the expansion of (\ref{bk}) yields

\bear
\Gamma_{0}^{a_{1}a_{2} a_{3}}[p_{1},\varepsilon_{1};p_{2},\varepsilon_{2};p_{3},\varepsilon_{3}]
 &=&(-ig)^{3}\mbox{tr}(T^{a_{1}} T^{a_{2}}T^{a_{3}})\int_{0}^{\infty} dT(4\pi T)^{-D/2}e^{-m^2 T}\non
 && \times\int_{0}^{T}d\tau_{1}\int_0^{\tau_{1}}d\tau_{2}\, (-i)^3 P_3 \,e^{(\cdot)}
\label{expand3point}
\ear
where

\bear
P_3 &=&
\dot{G}_{B1i}\varepsilon_{1}\cdot p_{i}\dot{G}_{B2j}\varepsilon_{2}\cdot p_{j}\dot{G}_{B3k}\varepsilon_{3}\cdot p_{k}
-\ddot{G}_{B12} \varepsilon_{1}\cdot\varepsilon_{2}\dot{G}_{B3k}\varepsilon_{3}\cdot p_{k}
\non
&& -\ddot{G}_{B13}\varepsilon_{1}\cdot\varepsilon_{3}\dot{G}_{B2j}\varepsilon_{2}\cdot p_{j}
-\ddot{G}_{B23}\varepsilon_{2}\cdot\varepsilon_{3}\dot{G}_{B1i}\varepsilon_{1}\cdot p_{i}\non 
\label{P3} 
\ear
and we have introduced the convention that  repeated indices $i,j,k,\ldots $ are to be summed from 1 to $N=3$. 
To remove, e.g., the term involving $\ddot G_{B12}\dot G_{B31}$ in the second term of $P_3$, we add the total derivative

\bear
-\frac{\partial}{\partial \tau_2}\Bigl(\dot{G}_{B12} \varepsilon_{1}\cdot\varepsilon_{2}\dot{G}_{B31}\varepsilon_{3}\cdot p_{1}
e^{(G_{B12}p_{1}\cdot p_{2}+G_{B13}p_{1}\cdot p_{3}+G_{23}p_{2}\cdot p_{3})}\Bigr) \, .
\label{add}
\ear
Adding five more similar total derivative terms removes all the $\ddot G_B$'s. Decomposing the
new integrand according to its ``cycle content'', 
$P_3$ gets replaced by  $Q_3=Q_3^3+Q_3^2$, where

\bear
Q_{3}^3&=&\dot{G}_{B12}\dot{G}_{B23}\dot{G}_{B31}Z_3(123)  \, ,\non
Q_{3}^2&=&\dot{G}_{B12}\dot{G}_{B21}Z_2(12)\dot{G}_{B3k}\varepsilon_{3}\cdot p_{k}+
\dot{G}_{B13}\dot{G}_{B31}Z_2(13)\dot{G}_{B2j}\varepsilon_{2}\cdot p_{j}
\non &&
+\dot{G}_{B23}\dot{G}_{B32}Z_2(23)\dot{G}_{B1i}\varepsilon_{1}\cdot p_{i} \, .
\non
\label{Q3}
\ear
Note that $Q^3_3$ contains a $\tau$-cycle of length three and $Q_3^2$ of length two, as indicated by the upper indices,
and that each $\tau$-cycle appears together with the corresponding ``Lorentz-cycle'', as advertised in the introduction.
The terms of $Q_3^2$ have, apart from the cycle, also a ``one-tail'', defined by \cite{strassler2}

\bear
T_1(a):= \varepsilon_a\cdot p_i \dot G_{Bai} \, .
\label{deftail}
\ear 
Although the form of the integrand reached in (\ref{Q3}) is already suitable for the application of the 
Bern-Kosower rules, it is natural to ask whether the polarization vectors appearing in the tails can also somehow be
completed to field strength tensors. 
Now in this three-point case there are already various chains of integrations-by-part that can be used to remove all the $\ddot G_{B}$'s,
but  if one assumes that the corresponding total derivative terms are added with constant coefficients (i.e., they involve no functions of 
momentum or polarization other than the ones already present in the term which one wishes to modify), then it is easy to convince oneself that they all
lead to the same $Q_3$ of (\ref{Q3}). This applies, in particular, to the ``old'' IBP procedure proposed in \cite{26}, where this $Q_3$ is obtained
by a chain of IBPs different from the above. 
Thus a more general type of IBPs is called for if one wishes to achieve this ``covariantization of the tails'', and in a companion paper \cite{91}
it will be shown how, using a more general type of total derivative terms with coefficients that {\it do} depend on 
momenta and polarizations, this can indeed be done for arbitrary $N$. 
Here we need not discuss this matter in more depth, since for $N=3$ the solution of this problem is still very simple.
Consider the first term in $Q_3^2$ above, eq. (\ref{Q3}).
Choose a momentum vector $r_3$ such that  $r_3\cdot p_3 \ne 0$, and add the total derivative

\bear
- \frac{r_3\cdot\varepsilon_3}{r_3\cdot p_3}Z_2(12)
\frac{\partial}{\partial\tau_3}\Bigl(\dot{G}_{B12}\dot{G}_{B21}e^{(\cdot)}\Bigr) \, .
\label{addtd}
\ear
The addition of this term to the first term in $Q_3^2$, and of similar terms to the second and third one,
transforms $Q_3^2$ into

\bear
R_3^2 &:=& \dot{G}_{B12}\dot{G}_{B21}Z_2(12)\dot{G}_{B3k}\frac{r_3\cdot f_3\cdot p_k}{r_3\cdot p_3}
+
\dot{G}_{B13}\dot{G}_{B31}Z_2(13)\dot{G}_{B2j}\frac{r_2\cdot f_2\cdot p_j}{r_2\cdot p_2}
\non
&& + \dot{G}_{B23}\dot{G}_{B32}Z_2(23)\dot{G}_{B1i}\frac{r_1\cdot f_1\cdot p_i}{r_1\cdot p_1}
\, .
\nonumber\\
\label{defR32}
\ear
Thus now all polarization vectors have been absorbed into tensors $f_i$, leading to manifest transversality.
This IBP procedure can be systematized to obtain closed-form integral 
representations of the Scalar and Spinor QED $N$ - photon
amplitudes that are manifestly gauge invariant at the integrand level \cite{91}.

Here, however, we are in the nonabelian case, where
the color-induced restriction of the parameter integrations to ordered sectors leads to the
appearance of boundary terms in the IBP \cite{strassler,strassler2}.
Let us look again at our total derivative term (\ref{add}). 
In the abelian case it would be integrated over the whole circle, and the result would be zero, since the worldline
Green's function $G_B(\tau_1,\tau_2)$ has the appropriate periodicity properties to make the two boundary terms
cancel. Here instead we find a nonzero result:

\bear
-\dot{G}_{B12} \varepsilon_{1}\cdot\varepsilon_{2}\dot{G}_{B31}\varepsilon_{3}\cdot p_{1}e^{(\cdot)}
\Big\vert_{\tau_2=\tau_3}^{\tau_2=\tau_1}
= 0 + \dot{G}_{B13} \varepsilon_{1}\cdot\varepsilon_{2}\dot{G}_{B31}\varepsilon_{3}\cdot p_{1}e^{G_{B13}p_{1}\cdot (p_{2}+p_{3})}
\, .
\nonumber\\
\label{bt}
\ear
Now, in the three-point case there are already two inequivalent orderings, say, $(123)$ and $(132)$; thus the full amplitude will
also have a part $\Gamma^{a_{1}a_{3} a_{2}}$ with color trace $\tr (T^{a_1}T^{a_3}T^{a_2})$, and the same total derivative term will
contribute to it a boundary term

\bear
-\dot{G}_{B12} \varepsilon_{1}\cdot\varepsilon_{2}\dot{G}_{B31}\varepsilon_{3}\cdot p_{1}e^{(\cdot)}
\Big\vert_{\tau_2=\tau_1}^{\tau_2=\tau_3}
= - \dot{G}_{B13} \varepsilon_{1}\cdot\varepsilon_{2}\dot{G}_{B31}\varepsilon_{3}\cdot p_{1}e^{G_{B13}p_{1}\cdot (p_{2}+p_{3})} - 0 \, .
\nonumber\\
\label{btother}
\ear
These two boundary terms would cancel in the abelian case, but now instead combine to produce a color commutator $\tr (T^{a_1}[T^{a_2},T^{a_3}])$. 
Moreover, among the other five similar total derivative terms needed to convert $P_3$ into $Q_3$ there is
one that differs from (\ref{add}) only by the interchange $2\leftrightarrow 3$. With some relabeling of integration variables, we can combine
the two boundary terms generated by that term with the two above to the structure

\bear
\tr (T^{a_1}[T^{a_2},T^{a_3}]) 
\varepsilon_3\cdot f_1\cdot\varepsilon_2
\dot{G}_{B12}\dot{G}_{B21}\,e^{G_{B12}p_{1}\cdot (p_{2}+p_{3})}  \, .
\label{structure}
\ear
Comparing with (\ref{Piscal}), we note that this term yields a parameter integral identical to the one 
of the two-gluon amplitude, except for the replacement of $p_2$ by $p_2+p_3$. In terms of the effective action,
from (\ref{defF}) and (\ref{structure}) 
its role is evidently to provide a piece needed to extend the ``abelian'' Maxwell term $\tr(f_{\mn}f^{\mn})$ to the
full nonabelian one $\tr(F_{\mn}F^{\mn})$. We will discuss this in more detail in section \ref{ea} below.

\no
To summarize so far, we can decompose the three-point amplitude for the scalar loop as 
(here and in the following we will often suppress the superscript ``$a_1a_2a_3$'')

\bear
\Gamma_{0} &=& 
\frac{g^3}{(4\pi)^{\frac{D}{2}}}(\Gamma_{0}^{3} 
+ \Gamma_{0}^{2} + \Gamma_{0}^{{\rm bt}})
\label{decomposescal}
\ear
where 

\bear
\Gamma_{0}^{3} &=& - \mbox{tr}(T^{a_{1}} T^{a_{2}}T^{a_{3}})\int_{0}^{\infty} \frac{dT}{T^{\frac{D}{2}}}e^{-m^2 T}\int_{0}^{T}d\tau_{1}\int_0^{\tau_{1}}d\tau_{2}\, Q_3^3 \,e^{(\cdot)}\non
&& -   \mbox{tr}(T^{a_{1}} T^{a_{3}}T^{a_{2}})\int_{0}^{\infty} \frac{dT}{T^{\frac{D}{2}}}e^{-m^2 T}\int_{0}^{T}d\tau_{1}\int_0^{\tau_{1}}d\tau_{3}\, Q_3^3
 \,e^{(\cdot)}\, ,\non
\Gamma_{0}^{2} &=& \Gamma_{0}^{3}(Q_3^3\to Q_3^2)\, , \non
\Gamma_{0}^{{\rm bt}} &=& \mbox{tr}(T^{a_{1}}[T^{a_{2}},T^{a_{3}}])
\int_{0}^{\infty} \frac{dT}{T^{\frac{D}{2}}}e^{-m^2 T}\int_{0}^{T}d\tau_{1}
\dot{G}_{B12}\dot{G}_{B21} \non
&&\times 
\Bigl\lbrack\varepsilon_3\cdot f_1\cdot\varepsilon_2
\,e^{G_{B12}p_{1}\cdot (p_{2}+p_{3})} +
\varepsilon_1\cdot f_2 \cdot\varepsilon_3
\,e^{G_{B12}p_{2}\cdot (p_{1}+p_{3})}
\nonumber\\ && \hspace{20pt}
 +
\varepsilon_2\cdot f_3\cdot\varepsilon_1
\,e^{G_{B12}p_{3}\cdot (p_{1}+p_{2})}
\Bigr\rbrack \non
\label{Gammas}
\ear
(here and in the following it is understood that always the last integration is eliminated by setting its integration variable
equal to zero; e.g., for the ordering $\tau_1>\tau_3>\tau_2$ we set $\tau_2=0$).
 
Alternatively, we can replace $Q_3^2$ by $R_3^2$ in the $\Gamma_{0}^2$ part, but then we have to also
add to $\Gamma_{0}^{{\rm bt}}$ a term $\tilde\Gamma_{0}^{{\rm bt}}$ containing the further boundary
contributions coming from the total derivative terms of the type (\ref{addtd}). Collecting those, one finds

\bear
\tilde\Gamma_{0}^{{\rm bt}} &=&- \half\mbox{tr}(T^{a_{1}}[T^{a_{2}},T^{a_{3}}])
\int_{0}^{\infty} \frac{dT}{T^{\frac{D}{2}}}e^{-m^2 T}\int_{0}^{T}d\tau_{1}
\dot{G}_{B12}\dot{G}_{B21} 
\nonumber\\
&& \hspace{-40pt}\times
\biggl\lbrace
\Bigl[{\rm tr}(f_1f_2)\rho_3-{\rm tr}(f_3f_1)\rho_2\Bigr]
\,e^{G_{B12}p_{1}\cdot (p_{2}+p_{3})} 
+
\Bigl[{\rm tr}(f_2f_3)\rho_1-{\rm tr}(f_1f_2)\rho_3\Bigr]
\,e^{G_{B12}p_{2}\cdot (p_{1}+p_{3})}
\nonumber\\ &&  \hspace{-20pt}
 +
\Bigl[{\rm tr}(f_3f_1)\rho_2-{\rm tr}(f_2f_3)\rho_1\Bigr]
\,e^{G_{B12}p_{3}\cdot (p_{1}+p_{2})}
\biggr\rbrace
\non
\label{Gammabtprime}
\ear
where we have now abbreviated $\rho_i:= r_i\cdot\varepsilon_i/r_i\cdot p_i$.

\section{The spinor loop case}
\label{section-spinor}
\renewcommand{\theequation}{3.\arabic{equation}}
\setcounter{equation}{0}

For the spinor loop case, it will be convenient to use the worldline super formalism 
\cite{rajeev,polyakovbook,andtse,strasslerthesis,15,sato2,18}.
In this formalism, one defines for each gluon leg a Grassmann variable $\theta_i$, 
$\theta^{2}_{i}=0$, and also considers the polarization vectors
$\varepsilon_i$ as being Grassmann. Thus all $\varepsilon_j^{\nu},\theta_k$, and $d\theta_l$
anticommute with each other. One further introduces the superderivative

\bear
D=\frac{\partial}{\partial\theta}-\theta\frac{\partial}{\partial\tau}
\label{defD}
\ear
and the super proper-time distance

\bear
\hat \tau_{ij}:= \tau_i - \tau_j +\theta_i\theta_j \, .
\label{defhattau}
\ear
Then the Bern-Kosower master formula can be generalized to the case of a Dirac fermion loop case as follows \cite{strasslerthesis,41}:

\begin{eqnarray}
\Gamma_{\half}^{a_{1}\dots a_{N}}
[p_1,\varepsilon_1;\ldots;p_N,\varepsilon_N]
&=&
-2
{(-ig)}^N
\mbox{tr}(T^{a_{1}}\dots T^{a_{N}})
{\dps\int_{0}^{\infty}}{dT\over T}
{(4\pi T)}^{-{D\over 2}}e^{-m^2T}
\non &&\times
\prod_{k=1}^N \int_0^T 
d\tau_k
\int
d\theta_k
 \delta\Bigl(\frac{\tau_{N}}{T}\Bigr)\vartheta(\hat\tau_{1N}) \prod_{l=1}^{N-1}\vartheta(\hat\tau_{l(l+1)})
 \nonumber\\&&
\hspace{-90pt}
\times
\exp\Biggl\lbrace
\sum_{i,j=1}^N
\Biggl\lbrack
\half\hat G_{ij} p_i\cdot p_j
+iD_i\hat G_{ij}\varepsilon_i\cdot p_j
+\half D_iD_j\hat G_{ij}\varepsilon_i\cdot\varepsilon_j\Biggr]
\Biggr\rbrace
\Biggl\vert_{\rm lin (\varepsilon_1 \ldots \varepsilon_N)} \, .
\nonumber\\
\label{supermaster}
\end{eqnarray}
\no
Here $\vartheta$ is the Heaviside step function and the Green's functions
$G_B$ and $G_F$ appear now combined into the super Green's function

\bear
\hat G(\tau_i,\theta_i;\tau_j,\theta_j)
\equiv G_B(\tau_i,\tau_j) +
\theta_i\theta_j G_F(\tau_i,\tau_j) \, .
\label{superpropagator}
\ear
The overall sign of (\ref{supermaster}) refers to the
standard ordering of the polarization vectors
$\varepsilon_1\varepsilon_2\ldots\varepsilon_N$.
Next, note that

\bear
\vartheta(\hat\tau_{ij}) = \vartheta(\tau_i-\tau_j)+\theta_i\theta_j\delta(\tau_i-\tau_j) \, .
\label{expandtheta}
\ear
The terms arising in the expansion of the spinor loop master formula (\ref{supermaster})
can be divided into three types: (i) terms that were there already for the scalar loop,
(ii) new terms not involving any of the delta functions appearing in (\ref{expandtheta}) 
and (iii) terms that do involve such
delta functions (only single delta function can appear up to the three-point level).
Concerning the type (ii) terms, those are known already from the abelian case, and it was shown in
\cite{41} by a direct combinatorial argument, starting from the abelian version of (\ref{supermaster}), 
that they can be taken into account correctly by the ``loop replacement rule'' eq. (\ref{subrule}). 
Terms of type (iii) are specific to the nonabelian case. They would cancel between adjacent ordered sectors in the abelian
case, but now produce color commutators, so that it is natural to think of them as a fermionic counterpart to the boundary terms
encountered in the scalar loop calculation. 

The loop replacement rule applies already in the two-point case, where the $\dot{G}_{B12}\dot{G}_{B21}$ appearing in
$Q_2$ has to be replaced by $\dot{G}_{B12}\dot{G}_{B21}-G_{F12}G_{F21}$.
Also a term of type (iii) appears at the two-point level, but it gives an integrand proportional to

\bear
\delta(\tau_1-\tau_2)G_F(\tau_1,\tau_2)\,\e^{G_B(\tau_1,\tau_2)}
\label{intvan}
\ear
that vanishes since $G_F(\tau,\tau) = 0$ by antisymmetry.
Taking the global normalization into account, the vacuum polarization function for the spinor loop case
becomes

\bear
\Pi_{\half}(p^2) &=& - 2 C( r)\frac{g^2}{(4\pi)^{D/2}}
\int_{0}^{\infty} \frac{dT}{T^{\frac{D}{2}}}e^{-m^2 T}
\non &&
\int_{0}^{T}d\tau_{1}\bigl(\dot{G}_{B12}\dot{G}_{B21}-G_{F12}G_{F21}\bigr) \,\e^{-G_{B12}p^2} \, .
\non
\label{Pispin}
\ear
In the three-point case, the effect of the type (ii) terms is to change each of the ``bulk terms'' $\Gamma^{2,3}_{0}$
of (\ref{Gammas}) to a corresponding $\Gamma^{2,3}_{\half}$ differing from its scalar loop counterpart by
a change of $Q_3^{2,3}$ to $\hat Q_3^{2,3}$, where

\bear
\hat Q_{3}^3&=&\bigl(\dot{G}_{B12}\dot{G}_{B23}\dot{G}_{B31} - G_{F12}G_{F23}G_{F31}\bigr)
Z_3(123) \, , \non
\hat Q_{3}^2&=&\bigl(\dot{G}_{B12}\dot{G}_{B21}-G_{F12}G_{F21}\bigr)Z_2(12)\dot{G}_{B3k}\varepsilon_{3}\cdot p_{k}+
\,\,{\rm two}\,\,{\rm permutations} \, .
\non
\label{superQ3}
\ear
But now  terms of type (iii) will really contribute. As mentioned above, those are similar to the boundary terms of the scalar
loop case, and a simple calculation shows that their role is precisely to make the ``loop replacement rule'' work for the
boundary terms, too. Thus for the total spinor loop contribution we find

\bear
\Gamma_{\half} &=& -2\frac{g^3}{(4\pi)^{\frac{D}{2}}}(\Gamma_{\half}^3 + \Gamma_{\half}^2 + \Gamma_{\half}^{{\rm bt}})
\label{decomposespin}
\ear
where

\bear
\Gamma_{\half}^3 &=&- \mbox{tr}(T^{a_{1}} T^{a_{2}}T^{a_{3}})\int_{0}^{\infty} 
\frac{dT}{T^{\frac{D}{2}}}e^{-m^2 T}\int_{0}^{T}d\tau_{1}\int_0^{\tau_{1}}d\tau_{2}\, \hat Q_3^3\vert_{\tau_3=0} \,e^{(\cdot)}\non
&& -   \mbox{tr}(T^{a_{1}} T^{a_{3}}T^{a_{2}})\int_{0}^{\infty}
\frac{dT}{T^{\frac{D}{2}}}e^{-m^2 T}\int_{0}^{T}d\tau_{1}\int_0^{\tau_{1}}d\tau_{3}\, \hat Q_3^3\vert_{\tau_2=0}  \,e^{(\cdot)}\, ,\non
\Gamma_{\half}^2 &=& \Gamma_{\half}^3(\hat Q_3^3\to \hat Q_3^2) \, ,\non
\Gamma_{\half}^{{\rm bt}} &=& \mbox{tr}(T^{a_{1}}[T^{a_{2}},T^{a_{3}}])
\int_{0}^{\infty} \frac{dT}{T^{\frac{D}{2}}}e^{-m^2 T}\int_{0}^{T}d\tau_{1}
\bigl(\dot{G}_{B12}\dot{G}_{B21}-G_{F12}G_{F21}\bigr)\non
&&\hspace{-25pt}\times
\Bigl\lbrack\varepsilon_3\cdot f_1\cdot\varepsilon_2
\,e^{G_{B12}p_{1}\cdot (p_{2}+p_{3})} +
\varepsilon_1\cdot f_2 \cdot\varepsilon_3
\,e^{G_{B12}p_{2}\cdot (p_{1}+p_{3})} +
\varepsilon_2\cdot f_3\cdot\varepsilon_1
\,e^{G_{B12}p_{3}\cdot (p_{1}+p_{2})}
\Bigr\rbrack\, .\non
\label{Gammasspin}
\ear
The alternative form of the scalar loop result, involving $R_3^2$ instead of $Q_3^2$ and the additional boundary contribution 
$\tilde\Gamma_{0}^{{\rm bt}}$,
can similarly be generalized to the spinor loop case by an application of the replacement rule (\ref{subrule}).

\section{The gluon loop case}
\label{section-gluon}
\renewcommand{\theequation}{4.\arabic{equation}}
\setcounter{equation}{0}

As was already mentioned, the case of the gluon loop is intrinsically more subtle,
because here one has the issue of gauge (in)dependence
not only for the background field but also for the loop particle. 
The preferred way of fixing the corresponding ambiguity for the three-vertex leads to the
``gauge-invariant vertex'', which obeys the simple Ward identity (\ref{ward}). 
This version of the vertex is generated by the BFM with Feynman
gauge for the quantum part, and it so happens that the only generalization of the
worldline path integral representation (\ref{avpi}) of the effective action to the gluon-loop 
case presently known is just based on the BFM with quantum Feynman gauge,
and thus the right starting point for a calculation of the ``gauge-invariant vertex''.
This representation was developed in \cite{strassler,18} in component fields, and reformulated
in terms of worldline superfields in \cite{sato2}. 
Although from the point of view of string theory it is a ``poor man's version'' of the Polyakov path integral
in the infinite string tension limit, it is perfectly adequate as far as the 1PI amplitudes are concerned.
It is also consistent  with full string theory in the sense that in the approach of \cite{dlmmr:plb351,dlmmr:npb469}, too,
the field theory limit of the off-shell continued string gluon amplitudes naturally leads to the 
Green's functions corresponding to the BFM with Feynman gauge. 

We will continue to take a user's approach here and proceed directly to the relevant master formula; the interested reader
may  consult \cite{41} for more details. This master formula for the (color-ordered) contribution
to the off-shell 1PI $N$-gluon  amplitude due to a gluon loop reads

\begin{eqnarray}
\Gamma_{\rm gluon}^{a_{1}\dots a_{N}}
[p_1,\varepsilon_1;\ldots;p_N,\varepsilon_N]
&=&
- \frac{(-ig)^N}{4}
\mbox{tr}(T^{a_{1}}\dots T^{a_{N}})
\lim_{C\to\infty}
{\dps\int_{0}^{\infty}}{dT\over T}
{(4\pi T)}^{-{D\over 2}}e^{-CT}
\nonumber\\&&\hspace{-120pt}
\times
\prod_{k=1}^N \int_0^T 
d\tau_k
\int
d\theta_k
\, \delta\Bigl(\frac{\tau_{N}}{T}\Bigr)
\vartheta(\hat\tau_{1N}) \prod_{l=1}^{N-1}\vartheta(\hat\tau_{l(l+1)})
 \sum_{p=P,A}
\sigma_p
Z_p
\nonumber\\&&\hspace{-130pt}\times
\exp\Biggl\lbrace
\sum_{i,j=1}^N
\Biggl\lbrack
\half\hat G_{p,ij}^C p_i\cdot p_j
+iD_i\hat G_{p,ij}^C\varepsilon_i\cdot p_j
+\half D_iD_j\hat G_{p,ij}^C\varepsilon_i\cdot\varepsilon_j\Biggr]
\Biggr\rbrace
\Biggl\vert_{\rm lin (\varepsilon_1 \ldots \varepsilon_N)}
\, .
\nonumber\\
\label{gluonmaster}
\end{eqnarray}
\no
Here the generators $T^a$ are now fixed to be in the adjoint representation. 
We have defined $\sigma_P=1$, $\sigma_A=-1$ (corresponding to periodic ($p=P$) and antiperiodic ($p=A$) boundary conditions
in the original path integral), and

\bear
Z_A&=&(2\cosh[CT/2])^4\, ,\nonumber\\
Z_P&=&(2\sinh[CT/2])^4\, ,\nonumber\\
\label{calcZAP}
\ear\no

\bear
\hat G^C_{P,A}
(\tau_1,\theta_1;\tau_2,\theta_2)
&=&
G_B(\tau_1,\tau_2)
+\theta_1\theta_2
{G}^C_{P,A}
(\tau_1,\tau_2) \, ,
\label{defGhatPA}
\ear
where

\bear\label{calcGchi}
{G}^C_P(\tau_1,\tau_2)&=&
2{\rm sign}(\tau_1-\tau_2)\frac{\sinh[C(\frac{T}{2}
-|\tau_1-\tau_2|)]}{\sinh[CT/2]} \, , \nonumber\\
{G}^C_A(\tau_1,\tau_2)&=&
2{\rm sign}(\tau_1-\tau_2)\frac{\cosh[C(\frac{T}{2}
-|\tau_1-\tau_2|)]}{\cosh[CT/2]} \, .
\nonumber\\
\label{defGAP}
\ear
The limit $C\to\infty$ and sum $\sum_{p=P,A}$ serve the purpose to remove unwanted degrees of freedom circulating in the loop.
Now, note that at fixed $C,p$ the gluon loop master formula (\ref{gluonmaster}) is isomorphic to the spinor loop  one (\ref{supermaster}).
For the two-point case, this formal analogy allows us to reuse (\ref{Pispin}) and write the vacuum polarization function due to a gluon loop in the form

\bear
\Pi_{\rm gluon}(p^2) &=&- \frac{1}{4} C( r)\frac{g^2}{(4\pi)^{D/2}}{\rm lim}_{C\to\infty}\sum_{p=P,A}\sigma_p
\int_{0}^{\infty} \frac{dT}{T^{\frac{D}{2}}}e^{-CT}Z_p
\non &&\times 
\int_{0}^{T}d\tau_{1}
\bigl(\dot{G}_{B12}\dot{G}_{B21}-G_{p12}^CG_{p21}^C\bigr) \,\e^{-G_{B12}p^2} \, .
\label{Piglu}
\ear
Similarly, in the three-point case the isomorphism implies that we can 
generalize the decomposition (\ref{decomposespin}) to

\bear
\Gamma_{\rm gluon} &=& -\frac{1}{4}\frac{g^3}{(4\pi)^{D/2}}\lim_{C\to\infty}
\sum_{p=P,A}\sigma_p\left(\Gamma_{\rm gluon}^3\bigl(C,p) 
+ \Gamma_{\rm gluon}^2(C,p) + \Gamma_{\rm gluon}^{{\rm bt}}(C,p)\right)
\nonumber\\
\label{decomposeglu}
\ear
where $\Gamma_{\rm gluon}^{(\cdot )}(C,p)$ differs from the corresponding $\Gamma_{\half}^{(\cdot )}$ in 
(\ref{Gammasspin}) only  by a replacement of $m^2$ by $C$, $G_{Fij}$ by $G^C_{p,ij}$, and the insertion of $Z_p$
under the $T$ integral.

It remains to analyze the limit $C\to\infty$ and the sum over boundary conditions; however, this has already been
done in complete generality in \cite{strassler,18,41}. 
For the integrands appearing in the two-point and three-point cases, the general rules found there give 

\bear
\lim_{C\to\infty}\,\e^{-CT}\sum_{p=P,A}\sigma_pZ_p &=& -8 \, ,\non
\lim_{C\to\infty}\,\e^{-CT}\sum_{p=P,A}\sigma_pZ_p G^C_{p,ij}G^C_{p,ji} &=& 16 \, ,  \non
\lim_{C\to\infty}\,\e^{-CT}\sum_{p=P,A}\sigma_pZ_p G^C_{p,12}G^C_{p,23}G^C_{p,31} &=& 16 \, . \non
\label{limC}
\ear
For the two-point case, this results in

\bear
\Pi_{\rm gluon}(p^2) &=& 2 C( r)\frac{g^2}{(4\pi)^{D/2}}
\int_{0}^{\infty} \frac{dT}{T^{\frac{D}{2}}}
\int_{0}^{T}d\tau_{1}
\bigl(\dot{G}_{B12}\dot{G}_{B21}+2\bigr) \,\e^{-G_{B12}p^2}
\, .
\non
\label{Piglufin}
\ear
For the three-point case we can write, from (\ref{decomposespin}), (\ref{Gammasspin}), (\ref{decomposeglu}) and (\ref{limC}),

\bear
\Gamma_{\rm gluon} &=& 2 \frac{g^3}{(4\pi)^{D/2}}(\Gamma_{\rm gluon}^3 + \Gamma_{\rm gluon}^2 + \Gamma_{\rm gluon}^{{\rm bt}})
\label{decomposegluon}
\ear
where, in terms of the spinor loop results of (\ref{Gammasspin}),

\bear
\Gamma_{\rm gluon}^3 &=& \Gamma_{\half}^3(G_{F12}G_{F23}G_{F31} \to - 2)\, ,\non
\Gamma_{\rm gluon}^2 &=&  \Gamma_{\half}^2(G_{F12}G_{F21} \to - 2)\, ,\non
\Gamma_{\rm gluon}^{{\rm bt}} &=& \Gamma_{\half}^{{\rm bt}}(G_{F12}G_{F21} \to - 2) \, .\non
\label{Gammasgluon}
\ear
However, we must not forget the ghost loop contribution, which is necessary for the subtraction of the
unphysical degrees of freedom of the gluon in the loop, and not contained in (\ref{gluonmaster}). 
This one is equal to the scalar loop contribution (\ref{Piscal}), but has to be taken with the opposite sign.
Thus e.g. for the two-point case we get the total spin-one contribution

\bear
\Pi_1(p^2) &\equiv &\Pi_{\rm gluon}(p^2) + \Pi_{\rm ghost}(p^2) \non
&=&
C( r)\frac{g^2}{(4\pi)^{D/2}}
\int_{0}^{\infty} \frac{dT}{T^{\frac{D}{2}}}
\int_{0}^{T}d\tau_{1}
\Bigl(\dot{G}_{B12}\dot{G}_{B21}+4\Bigr) \,\e^{-G_{B12}p^2}
\, .
\non
\label{Pi1}
\ear
Finally, in the gluon loop case, too, we have the option of using 
$R_3^2$ instead of $Q_3^2$ in the three-point vertex, 
with an additional boundary contribution $\tilde\Gamma_{\rm gluon}^{{\rm bt}}$, and it is easy to check that
this form of the result still relates to the corresponding one for the spinor loop result by  (\ref{Gammasgluon}).

\section{Summary}
\label{summary}
\renewcommand{\theequation}{5.\arabic{equation}}
\setcounter{equation}{0}

We will now summarize our results for the scalar, spinor, and gluon loop cases. For easy comparison,
here we also rewrite the multiple $\tau_i$ - integrals in terms of the more standard Feynman/Schwinger parameters $\alpha_i$.
First, as usual we rescale $\tau_i = Tu_i$, after which the $T$ integral can be done trivially. 
Then in the two-point case we set $u_2=0$ and change from $u_1$ to $\alpha$, 
and in the three-point case we set $u_3=0$ and change from $u_1,u_2$ to $\alpha_1,\alpha_2,\alpha_3$ via  

\bear
u_1 &=& {\alpha}_2+{\alpha}_3 \, ,\non
u_2 &=& {\alpha}_3 \, , \non
\label{trafotaua}
\ear
with ${\alpha}_1+{\alpha}_2+ {\alpha}_3= 1$. 
Only six different parameter integrals appear:

\bear
I_{\rm 2pt,B}^D(p^2) &=& \int_0^1d{\alpha}\frac{(1-2{\alpha})^2}{\bigl( m^2 + {\alpha}(1-{\alpha})p^2\bigr)^{2-\frac{D}{2}}} 
\, ,\non
I_{{\rm 2pt},F}^D(p^2) &=& \int_0^1d{\alpha}\frac{1}{\bigl( m^2 + {\alpha}(1-{\alpha})p^2\bigr)^{2-\frac{D}{2}}}
\, ,\non
I_{3,B}^D(p_1^2,p_2^2,p_3^2) &=& \int_0^1d{\alpha}_1d{\alpha}_2d{\alpha}_3\delta(1-{\alpha}_1-{\alpha}_2-{\alpha}_3)
\frac{(1-2{\alpha}_1)(1-2{\alpha}_2)(1-2{\alpha}_3)}{\Bigl( m^2 + {\alpha}_1{\alpha}_2p_1^2+{\alpha}_2{\alpha}_3p_2^2+{\alpha}_1{\alpha}_3p_3^2\Bigr)^{3-\frac{D}{2}}}
\, ,\non
I_{3,F}^D(p_1^2,p_2^2,p_3^2) &=& -\int_0^1d{\alpha}_1d{\alpha}_2d{\alpha}_3\delta(1-{\alpha}_1-{\alpha}_2-{\alpha}_3)
\frac{1}{\Bigl( m^2 + {\alpha}_1{\alpha}_2p_1^2+{\alpha}_2{\alpha}_3p_2^2+{\alpha}_1{\alpha}_3p_3^2\Bigr)^{3-\frac{D}{2}}}
\, ,\non
I_{2,B}^D(p_1^2,p_2^2,p_3^2) &=& \int_0^1d{\alpha}_1d{\alpha}_2d{\alpha}_3\delta(1-{\alpha}_1-{\alpha}_2-{\alpha}_3)
\frac{(1-2{\alpha}_2)^2(1-2{\alpha}_1)}{\Bigl( m^2 + {\alpha}_1{\alpha}_2p_1^2+{\alpha}_2{\alpha}_3p_2^2+{\alpha}_1{\alpha}_3p_3^2\Bigr)^{3-\frac{D}{2}}}
\, ,\non
I_{2,F}^D(p_1^2,p_2^2,p_3^2) &=& \int_0^1d{\alpha}_1d{\alpha}_2d{\alpha}_3\delta(1-{\alpha}_1-{\alpha}_2-{\alpha}_3)
\frac{1-2{\alpha}_1}{\Bigl( m^2 + {\alpha}_1{\alpha}_2p_1^2+{\alpha}_2{\alpha}_3p_2^2+{\alpha}_1{\alpha}_3p_3^2\Bigr)^{3-\frac{D}{2}}}\, . \non
\label{I}
\ear
Our results for the vacuum polarization tensors (\ref{Piscal}), (\ref{Pispin}), (\ref{Pi1}) can then be summarized as

\bear
\Pi_0(p^2) &=& - C(r)\frac{g^2}{(4\pi)^{D/2}}\Gamma\left(2-\frac{D}{2}\right)I^D_{{\rm 2pt}, B}(p^2)\, , \nonumber\\
\Pi_{\frac{1}{2}}(p^2) &=& -2 \Pi_0(p^2)\Bigl(I^D_{{\rm 2pt}, B} \to I^D_{{\rm 2pt}, B} - I^D_{{\rm 2pt},F}\Bigr)\, ,\nonumber\\
\Pi_{1}(p^2) &=& \Pi_0(p^2)\Bigl(I^D_{{\rm 2pt}, B} \to I^D_{{\rm 2pt}, B} - 4I^D_{{\rm 2pt},F}\Bigr)\, . \nonumber\\
\label{Pis}
\ear
Here it is understood that the formula for $\Pi_1$ refers to the adjoint representation and to the massless case.

\no
For the three-point case, too, we can unify our results as follows:

\bear
\Gamma_s &=& d_s\frac{g^3}{(4\pi)^{\frac{D}{2}}}(\Gamma_{\rm s}^3 + \Gamma_{\rm s}^2 + \Gamma_{\rm s}^{{\rm bt}}) \, .
\label{decomposeallfin}
\ear
Here $d_0=d_1=1,d_{\half} = -2$ and

\bear
\Gamma_0^3 &=&-\Gamma\left(3-\frac{D}{2}\right)\mbox{tr}(T^{a_{1}} T^{a_{2}}T^{a_{3}})\tr (f_1f_2f_3)
I^D_{3,B}(p_1^2,p_2^2,p_3^2)
+ (a_2\leftrightarrow a_3, f_2\leftrightarrow f_3, p_2 \leftrightarrow p_3)
\, ,\non
\Gamma_0^2 &=& \half\Gamma\left(3-\frac{D}{2}\right)\mbox{tr}(T^{a_{1}} T^{a_{2}}T^{a_{3}})
\Bigl\lbrack
\tr (f_1f_2)\bigl(\varepsilon_3\cdot p_1 I^D_{2,B}(p_1^2,p_2^2,p_3^2)-\varepsilon_3\cdot p_2 I^D_{2,B}(p_2^2,p_1^2,p_3^2)\bigr)\non
&&\hspace{130pt}+ \tr (f_2f_3)\bigl(\varepsilon_1\cdot p_2 I^D_{2,B}(p_2^2,p_3^2,p_1^2)-\varepsilon_1\cdot p_3 I^D_{2,B}(p_3^2,p_2^2,p_1^2)\bigr)\non
&&\hspace{130pt} + \tr (f_3f_1)\bigl(\varepsilon_2\cdot p_3 I^D_{2,B}(p_3^2,p_1^2,p_2^2)-\varepsilon_2\cdot p_1 I^D_{2,B}(p_1^2,p_3^2,p_2^2)\bigr)\Bigr\rbrack\non
&&+ (a_2\leftrightarrow a_3, f_2\leftrightarrow f_3, \varepsilon_2\leftrightarrow \varepsilon_3, p_2 \leftrightarrow p_3)
\, ,\non
\Gamma_0^{{\rm bt}} &=& -\Gamma\left(2-\frac{D}{2}\right)\mbox{tr}(T^{a_{1}}[T^{a_{2}},T^{a_{3}}])
\Bigl\lbrack\varepsilon_3 \cdot f_1\cdot\varepsilon_2 I^D_{{{\rm 2pt}},B}(p_1^2)+
\varepsilon_1 \cdot f_2\cdot \varepsilon_3 I^D_{{{\rm 2pt}},B}(p_2^2)
\non&& \hspace{160pt}
+
\varepsilon_2 \cdot f_3\cdot\varepsilon_1 I^D_{{{\rm 2pt}},B}(p_3^2)
\Bigr\rbrack \, ,\non
\label{Gammas0fin}
\ear

\bear
\Gamma_{\frac{1}{2}}^3 &=& - \Gamma\left(3-\frac{D}{2}\right)\mbox{tr}(T^{a_{1}} T^{a_{2}}T^{a_{3}})\tr (f_1f_2f_3)
\bigl(I^D_{3,B}(p_1^2,p_2^2,p_3^2) - I^D_{3,F}(p_1^2,p_2^2,p_3^2)\bigr)
+\, (2\leftrightarrow 3)
\, ,\non
\Gamma_{\frac{1}{2}}^2 &=&
\half\Gamma\left(3-\frac{D}{2}\right)\mbox{tr}(T^{a_{1}} T^{a_{2}}T^{a_{3}})
\Bigl\lbrack
\tr (f_1f_2)\Bigl(\varepsilon_3\cdot p_1 \bigl(I^D_{2,B}(p_1^2,p_2^2,p_3^2)-I^D_{2,F}(p_1^2,p_2^2,p_3^2)\bigr) \non
&&\hspace{130pt}
-\varepsilon_3\cdot p_2 \bigl( I^D_{2,B}(p_2^2,p_1^2,p_3^2)-I^D_{2,F}(p_2^2,p_1^2,p_3^2)\bigr)\Bigr)
+ \,\, {\rm 2 \,\, perm.}
\Bigr\rbrack\non
&&+  (2\leftrightarrow 3)
\, ,\non
\Gamma_{\frac{1}{2}}^{{\rm bt}} &=& 
-\Gamma\left(2-\frac{D}{2}\right)\mbox{tr}(T^{a_{1}}[T^{a_{2}},T^{a_{3}}])
\Bigl\lbrack\varepsilon_3 \cdot f_1\cdot\varepsilon_2 \bigl(I^D_{{{\rm 2pt}},B}(p_1^2)-I^D_{{{\rm 2pt}},F}(p_1^2)\bigr) + \,\, {\rm 2 \,\, perm.}
\Bigr\rbrack\, ,\non
\label{Gammashalffin}
\ear

\bear
\Gamma_{1}^3 &=& - \Gamma\left(3-\frac{D}{2}\right)\mbox{tr}(T^{a_{1}} T^{a_{2}}T^{a_{3}})\tr (f_1f_2f_3)
\bigl(I^D_{3,B}(p_1^2,p_2^2,p_3^2) -4 I^D_{3,F}(p_1^2,p_2^2,p_3^2)\bigr)
+ (2\leftrightarrow 3)
\, ,\non
\Gamma_{1}^2 &=&
\half\Gamma\left(3-\frac{D}{2}\right)\mbox{tr}(T^{a_{1}} T^{a_{2}}T^{a_{3}})
\Bigl\lbrack
\tr (f_1f_2)\Bigl(\varepsilon_3\cdot p_1 \bigl(I^D_{2,B}(p_1^2,p_2^2,p_3^2)-4I^D_{2,F}(p_1^2,p_2^2,p_3^2)\bigr) \non
&&\hspace{130pt}
-\varepsilon_3\cdot p_2 \bigl( I^D_{2,B}(p_2^2,p_1^2,p_3^2)-4I^D_{2,F}(p_2^2,p_1^2,p_3^2)\bigr)\Bigr)
+ \,\, {\rm 2 \,\, perm.}
\Bigr\rbrack\non
&&+  (2\leftrightarrow 3)
\, , \non
\Gamma_{1}^{{\rm bt}} &=& 
-\Gamma\left(2-\frac{D}{2}\right)\mbox{tr}(T^{a_{1}}[T^{a_{2}},T^{a_{3}}])
\Bigl\lbrack\varepsilon_3 \cdot f_1\cdot\varepsilon_2 \bigl(I^D_{{{\rm 2pt}},B}(p_1^2)-4I^D_{{{\rm 2pt}},F}(p_1^2)\bigr) + \,\, {\rm 2 \,\, perm.}
\Bigr\rbrack
\, .\non
\label{Gammas1fin}
\ear
The parameter integrals at the three-point level are already highly nontrivial, and we refer the reader to
\cite{daossa-3gluonDm}, and refs. therein, for methods for their evaluation. 

We note that the ``loop replacement rules'' now have, for both the two-point and three-point cases, assumed the form

\bear
\Gamma_{\frac{1}{2}}^{(\cdot)} &=& \Gamma_{0}^{(\cdot)}\Bigl(I^D_{(\cdot),B} \to I^D_{(\cdot),B} - I^D_{(\cdot),F}\Bigr)\, ,
 \label{replacespinor}\\
 \Gamma_{1}^{(\cdot)} &=& \Gamma_{0}^{(\cdot)}\Bigl(I^D_{(\cdot),B} \to I^D_{(\cdot),B} - 4 I^D_{(\cdot),F}\Bigr)\, .
 \label{replacegluon}
\ear
Further, it can be easily checked that, for the bulk terms $\Gamma_{(\cdot)}^{3,2}$, the terms with the interchange $(2\leftrightarrow 3)$ just provide
the other half of a color commutator $[T^{a_2},T^{a_3}]$, so that for them, as for the boundary terms, the color structure factors out in
a $\mbox{tr}(T^{a_{1}}[T^{a_{2}},T^{a_{3}}])$. Therefore we can now use

\bear
\mbox{tr}(T^{a_{1}}[T^{a_{2}},T^{a_{3}}]) = iC(r)f^{a_1a_2a_3}
\label{trtof}
\ear
to get the expected proportionality to $f^{a_1a_2a_3}$. Thus we can write

\bear
\Gamma_s^{a_1a_2a_3} &=& d_s\frac{g^3}{(4\pi)^{\frac{D}{2}}}\mbox{tr}(T^{a_{1}}[T^{a_{2}},T^{a_{3}}])(\gamma_{\rm s}^3 + \gamma_{\rm s}^2 + \gamma_{\rm s}^{{\rm bt}})\non
 &=&if^{a_1a_2a_3}C(r) d_s\frac{g^3}{(4\pi)^{\frac{D}{2}}}
 (\gamma_{\rm s}^3 + \gamma_{\rm s}^2 + \gamma_{\rm s}^{{\rm bt}})\non
\label{decomposeall}
\ear
with

\bear
\gamma_0^3 &=&  -\Gamma\left(3-\frac{D}{2}\right)\tr (f_1f_2f_3)
I^D_{3,B}(p_1^2,p_2^2,p_3^2)
\, ,\non
\gamma_0^2 &=& \half\Gamma\left(3-\frac{D}{2}\right)\Bigl\lbrack
\tr (f_1f_2)\Bigl(\varepsilon_3\cdot p_1 I^D_{2,B}(p_1^2,p_2^2,p_3^2)-\varepsilon_3\cdot p_2 I^D_{2,B}(p_2^2,p_1^2,p_3^2)\Bigr)\non
&&\hspace{56pt}+ \tr (f_2f_3)\Bigl(\varepsilon_1\cdot p_2 I^D_{2,B}(p_2^2,p_3^2,p_1^2)-\varepsilon_1\cdot p_3 I^D_{2,B}(p_3^2,p_2^2,p_1^2)\Bigr)\non
&&\hspace{56pt} + \tr (f_3f_1)\Bigl(\varepsilon_2\cdot p_3 I^D_{2,B}(p_3^2,p_1^2,p_2^2)-\varepsilon_2\cdot p_1 I^D_{2,B}(p_1^2,p_3^2,p_2^2)\Bigr)\Bigr\rbrack\, ,\non
\gamma_0^{{\rm bt}} &=& -\Gamma\left(2-\frac{D}{2}\right)
\Bigl\lbrack\varepsilon_3 \cdot f_1\cdot\varepsilon_2 I^D_{{{\rm 2pt}},B}(p_1^2)+
\varepsilon_1 \cdot f_2\cdot \varepsilon_3 I^D_{{{\rm 2pt}},B}(p_2^2)+
\varepsilon_2 \cdot f_3\cdot\varepsilon_1 I^D_{{{\rm 2pt}},B}(p_3^2)
\Bigr\rbrack\non
\label{gammas0fin}
\ear
and the $\gamma_{\frac{1}{2},1}^{(\cdot)}$'s obtained from the $\gamma_0^{(\cdot)}$'s by the rule (\ref{replacespinor}) resp. (\ref{replacegluon}). 

\no
In the version where $Q_3^2$ is traded for $R_3^2$, $\gamma_0^2$ gets replaced by 

\bear
\tilde\gamma_0^{2} &=& \half\Gamma\left(3-\frac{D}{2}\right)\Bigl\lbrace
\tr (f_1f_2)\Bigl[\frac{r_3\cdot f_3\cdot p_1}{r_3\cdot p_3} I^D_{2,B}(p_1^2,p_2^2,p_3^2)-
\frac{r_3\cdot f_3\cdot p_2}{r_3\cdot p_3} I^D_{2,B}(p_2^2,p_1^2,p_3^2)\Bigr]
+ \, 2 \,{\rm perm.} \Bigr\rbrace\non
\label{gamma02prime}
\ear
and one has the additional boundary contribution

\bear
\tilde\gamma_0^{{\rm bt}} &=& \half\Gamma\left(2-\frac{D}{2}\right)
\Bigl\lbrace
\Bigl[{\rm tr}(f_1f_2)\rho_3-{\rm tr}(f_3f_1)\rho_2\Bigr] I^D_{{{\rm 2pt}},B}(p_1^2)+\, 2 \,{\rm perm.} 
\Bigr\rbrace \, .\non
\label{gammabtprimefin}
\ear
The rules (\ref{replacespinor}) and (\ref{replacegluon}) continue to hold.

\section{Comparison with previous results}
\label{comp}
\renewcommand{\theequation}{6.\arabic{equation}}
\setcounter{equation}{0}

We now study the connection between our results for the three-gluon vertex and previous work.
Since our treatment of the gluon loop case is equivalent to the use of the BFM
with quantum Feynman gauge, we expect the Binger-Brodsky relation (\ref{relsusy}) to hold; and indeed
this relation here follows immediately from the replacement rules (\ref{replacespinor}) and (\ref{replacegluon}).
(Similarly we can use (\ref{Pis}) to verify the vanishing of the gluon propagator in $N=4$ SYM theory.)

For the same reason, the QED-like Ward identity (\ref{ward}) should be fulfilled not only for the scalar and spinor, but also
for the gluon loop case. Here it is advantageous to use the version where $R_3^2$ is used instead of $Q_3^2$. Since
$R_3^2$ is transversal, the Ward identity then involves only the boundary terms $\gamma_0^{{\rm bt}},\tilde\gamma_0^{{\rm bt}}$, 
and can be easily verified using (\ref{Pis}), (\ref{gammas0fin}), and (\ref{gammabtprimefin}).

Next, we proceed to the less straightforward task of relating our representation to the Ball-Chiu decomposition. 
As usual we start with the scalar case.
Comparing  our final result (\ref{decomposeall}),(\ref{gammas0fin}) with (\ref{B.C}), (\ref{vertexcontracted}) we first note that
$T_{H}={\rm tr}(f_{1}f_{2}f_{3})$. Thus we must identify

\bear
H(p_1^2,p_2^2,p_3^2)&=&
C(r)\frac{d_0g^2}{(4\pi)^{D/2}}\Gamma\left(3-\frac{D}{2}\right) I^{D}_{3,B} (p_1^{2},p_2^{2},p_3^{2})
\label{identifyH}
\ear
which is indeed totally symmetric in its arguments.

Further, it is also easy to 
recognize the functions $A$ and $B$ functions as symmetric and antisymmetric
combinations of the functions contained in $\gamma_{0}^{{\rm bt}}$: 

\bear
A (p_1^2,p_2^2;p_3^2)&=&
C(r)\frac{d_0g^2}{2(4\pi)^{D/2}}
 \Gamma\left(2-\frac{D}{2}\right)\Big[I^{D}_{{{\rm 2pt}, B}}(p_1^2)+I^{D}_{{{\rm 2pt}, B}}(p_2^2)\Bigr]\, ,\non
T_{A}&=&\varepsilon_{1}\cdot\varepsilon_{2}(p_{1}\cdot\varepsilon_{3}-p_{2}\cdot\varepsilon_{3})\non
\label{A}
\ear
and
\bear
B(p_1^2,p_2^2;p_3^2)&=&
C(r)\frac{d_0g^2}{2(4\pi)^{D/2}}
\Gamma\left(2-\frac{D}{2}\right)\Bigl[I^{D}_{{{\rm 2pt}, B}}(p_1^2)-I^{D}_{{{\rm 2pt}, B}}(p_2^2)\Bigr]\, ,\non
T_{B}&=&\varepsilon_{1}\cdot\varepsilon_{2}(p_{1}\cdot\varepsilon_{3}+p_{2}\cdot\varepsilon_{3})\, .\non
\label{B}
\ear
Coming to the structure $F$, the fact that $T_F$ is transversal suggests that we should again use the transversal structure $\tilde\gamma_0^{2}$ rather than
$\gamma_0^2$; the question is, how to choose the still undetermined vectors $r_i$? By inspection one finds that, with the cyclic choice
$r_1=p_2-p_3,r_2=p_3-p_1,r_3=p_1-p_2, $ and using
the antisymmetry of $f_i$, e.g. the first term in braces in (\ref{gamma02prime}) turns into 

\bear
{\rm tr}(f_1f_2) \frac{p_1\cdot f_3\cdot p_2}{(p_1-p_2)\cdot p_3}  \Bigl[I^D_{2,B}(p_1^2,p_2^2,p_3^2)-I^D_{2,B}(p_2^2,p_1^2,p_3^2)\Bigr]\, .
\label{TFinspect}
\ear

\no
Noting that

\bear
T_F = \half  {\rm tr}(f_1f_2) p_1\cdot f_3\cdot p_2
\label{idTF}
\ear
we are led to set

\bear
F(p_1^2,p_2^2;p_3^2)&=&
C(r)\frac{d_0g^2}{(4\pi)^{D/2}}\Gamma\left(3-\frac{D}{2}\right) 
\frac{I_{2,B}^{D}(p_1^2,p_2^2,p_3^2)-I_{2,B}^{D}(p_2^2,p_1^2,p_3^2)}{p_1^2-p_2^2}\non
\label{F}
\ear
where we have also used momentum conservation to rewrite 

\bear
(p_1-p_2)\cdot p_3 = p_2^2-p_1^2 \, .
\label{usemomcons}
\ear
Thus the remaining structure $C$ must match $\tilde\gamma_0^{{\rm bt}}$, and indeed one has

\bear
T_C = \half {\rm tr}(f_1f_2)(p_1-p_2)\cdot \varepsilon_3
= - \half {\rm tr}(f_1f_2)\rho_3 (p_1^2-p_2^2)
\label{idTC}
\ear
leading to the identification

\bear
C(p_1^2,p_2^2;p_3^2)&=&
C(r)\frac{d_0g^2}{(4\pi)^{D/2}}\Gamma\left(2-\frac{D}{2}\right) 
\frac{I^{D}_{{{\rm 2pt}, B}}(p_1^2)-I^{D}_{{{\rm 2pt}, B}}(p_2^2)}{p_1^2-p_2^2}\, .\non
\label{C}
\ear
Note that $F$ and $C$ are indeed symmetric functions in the first two arguments,
and that $C$ is actually independent of $p_3^2$. Also, $A$ is the only one of the
functions having a UV divergence (since in $B$ the 
expression in square brackets is $O(\epsilon)$),
and $B$ and $C$ are simply related by \cite{daossa-3gluonDm}

\bear
2B = (p_1^2-p_2^2)C \, .
\label{relBC}
\ear
Passing from the scalar to the spinor and gluon loop cases using 
(\ref{replacespinor}) and (\ref{replacegluon}) will obviously not change
anything essential in this analysis. 

For the (massive) spinor loop case we have also verified the above correspondences explicitly,
using the formulas for the functions $A$ to $H$ given in \cite{daossa-3gluonDm}
(to be precise, we have done this check for $A,B,C$ with arbitrary momentum, for $F$ 
specializing to $p_3^2=0$ and for $H$ specializing to $p_1^2=p_2^2=0$. This provides also a
check on the much more involved calculations of \cite{daossa-3gluonDm}).

\section{Comparison with the effective action}
\label{ea}
\renewcommand{\theequation}{7.\arabic{equation}}
\setcounter{equation}{0}

It will be instructive to compare our results for the three-point amplitude
with the low-energy expansion of the one-loop QCD effective action induced by a
loop particle of mass $m$. 
The general form of this expansion is 

\begin{equation}
\Gamma_{\rm 0}[F] = \int_0^\infty \!{dT\over T} \; 
\frac{{\rm e}^{-m^2 T}}{(4\pi T)^{D/2}} \; 
{\rm tr} \; \int \! dx_0 \; \sum_{n=2}^{\infty} \; 
\frac{(-T)^n}{n!} \; O_n[F] \,,
\label{LEE}
\end{equation}\no
where $O_n(F)$ is a Lorentz and gauge-invariant expression of mass dimension $2n$.
For the scalar loop, in \cite{23,25} this expansion was obtained to order $O(T^6)$.
To see the relation with our form factor decomposition, it will be sufficient
to consider the $n=2$ and $n=3$ terms:

\begin{eqnarray}
O_2 &=& - {1\over 6}g^2 F_{\mu\nu} F_{\mu\nu} \, ,\nonumber\\
O_3 &=& 
          - {2\over 15} \,i g^3\,\Fkala\Flamu\Fmuka
          - {1\over 20}g^2 D_{\lambda}F_{\mu\nu}D^{\lambda}F^{\mu\nu} \, .
          \nonumber\\
          \label{O23}
\end{eqnarray}
\no
Here changing from the scalar to the spinor or gluon loop will change only the coefficients
in the expansion (\ref{LEE}), not its structure. Comparing with, e.g., (\ref{gammas0fin})
we easily recognize the correspondences 

\begin{eqnarray}
&&\gamma_{(\cdot)}^3\leftrightarrow F_{\kappa}^{~\lambda}F_{\lambda}^{~\mu}F_{\mu}^{~\kappa}
=f_{\kappa}^\lambda f_{\lambda}^\mu f_{\mu}^\kappa+{\rm higher~point~terms}\, ,\nonumber\\
&&\gamma_{(\cdot)}^2\leftrightarrow (\partial+ig\underbrace {A)F(\partial}+igA)F\, ,\nonumber\\
&&\gamma_{(\cdot)}^{{\rm bt}}\leftrightarrow(f+ig\underbrace{[A,A])(f}+ig[A,A])\, .\nonumber\\
\label{comparison}
\end{eqnarray}
Thus all the pieces of our form factor decomposition have a simple meaning in terms of the effective action. Note that commutator terms
are always generated by boundary terms in the IBP, and that our three-point results allow us to predict certain terms in the higher-point
gluon amplitudes using the knowledge that any ``abelian'' field strength tensor in the nonabelian effective action must appear as part
of the full nonabelian field strength tensor including the commutator term. Note also that the tensor structure multiplying the function $S$
in the Ball-Chiu decomposition does not correspond to anything in the expansion (\ref{LEE}), which is one way of understanding why $S$ turned out to be zero in
the calculations of \cite{balchi2,daossa-3gluonDm}. Since the structure of the effective action is independent of the loop order,
this observation allows us also to predict that the vanishing of $S$ is not a one-loop accident, and will be found to persist at higher loop orders.

\section{Conclusions}
\label{conc}
\renewcommand{\theequation}{8.\arabic{equation}}
\setcounter{equation}{0}

We have recalculated here the one-loop QCD three-gluon vertex for the scalar, fermion and gluon
loop cases in a unifying way, achieving a compact result involving only six different parameter
integrals. We have established the precise relation of this result to the standard Ball-Chiu
decomposition, and also verified this relation for the massive spinor loop case using the explicit
results of \cite{daossa-3gluonDm}. 

As was mentioned already in the introduction, even in a four-dimensional calculation
the use of the vertex as a building block for higher-loop calculations
will in most cases make it necessary to know its $D$ - dimensional continuation.
For that reason we have have kept the full $D$ - dependence as much as possible. In fact, our
result for the scalar loop is complete in this sense and holds for arbitrary $D$.
And also in the spinor loop case the only place where we have used $D=4$ is in the normalization
of the path integral, which however corresponds to the usual fixing of ${\rm tr}_{\gamma}\Eins = 4$
in dimensional regularization. It is only in the gluon loop case where $D=4$ has been used in a nontrivial
way, namely already in the derivation of (\ref{gluonmaster}). Here a true extension to other spacetime
dimensions would require some more work. For the purpose of dimensional regularization it is, however,
sufficient to note that our result for the gluon loop case corresponds to the dimensional reduction variant
of dimensional regularization proposed in \cite{berkos:npb379}.

In this calculation, three main advantages of our approach have emerged.

First, the IBP procedure generates the standard transversality-based Ball-Chiu decomposition of the
vertex almost automatically, bypassing the usual tedious analysis of the nonabelian Ward identities.
Let us recapitulate how this happens: for the bulk terms, in the IBP all polarization vectors
get absorbed into ``abelian'' field strength tensors $f_i^{\mu\nu}$, and thus become transversal.
In the abelian case, there would be no boundary contributions, and one would have achieved manifest
transversality at the integrand level. In the nonabelian case there are boundary terms, and those
combine into commutator terms that carry all nontransversality, and generally contribute to the covariantization
of some lower-point bulk term. 

Second, this emergence of field strength tensors in the IBP allows one to maintain a close
relation between the momentum space amplitudes and the low-energy effective action, and thus to profit 
from the superior organization of the latter with respect to gauge invariance. 
This has led us to predict that the vanishing of the coefficient function $S$ of the Ball-Chiu
decomposition will be found to persist beyond one-loop.

Third, the integrands of the spinor and gluon loop contributions can be obtained from the scalar loop one trivially
using the off-shell extended Bern-Kosower ``loop replacement rules'' (\ref{replacespinor}) and (\ref{replacegluon}).
The gluon loop result corresponds to a field theory calculation in the BFM
with quantum Feynman gauge, and thus to the preferred ``gauge-invariant vertex'' 
which fulfills the simple Ward identity (\ref{ward}) and the SUSY-related identity (\ref{relsusy}).
The latter here appears as a simple consequence of the replacement rules and thus relates to 
worldline SUSY. 

Concerning the last point, in the calculation presented here we have verified, rather than 
assumed, the validity of this off-shell extension.
Had we taken the validity of the replacement rules off-shell for granted from the beginning, 
our method of calculating the three-point vertex would have been even much more efficient;
in fact, incomparably more efficient than the combined effort of \cite{balchi2,daossa-3gluonDm,binbro}
that was necessary to arrive at an explicit result for the scalar, spinor and gluon loop
contributions to the three-gluon vertex with standard field theory methods.
Before applying it to higher-point vertices, it will thus be important to show 
the validity of this off-shell extension in general. With hindsight, this can be done as follows:
it is sufficient to show the validity of the replacement rules for the effective action.
Let us consider the spinor loop case first. Here it was shown in \cite{41} 
that the replacement rule in the abelian 
case holds for the off-shell $N$ - photon amplitudes. Thus it holds also for the abelian
effective action. The nonabelian effective action in the low-energy expansion can be decomposed into
terms that are Lorentz scalars built from covariant derivatives and field strength tensors.  Each such term in
general will, after Fourier transformation, contribute to momentum space functions 
with various different numbers of legs;  
e.g., the term $\tr (D_{\mu}F_{\alpha\beta}D^{\mu}F^{\alpha\beta})$ will contribute to the $N$ - point functions with $N$ between two and six.
Generally, each such term in the nonabelian effective Lagrangian has a ``core'' term, which has a counterpart already in the abelian
case (in the example this would be $\partial_{\mu}f_{\alpha\beta}\partial^{\mu}f^{\alpha\beta}$) and a number of
``covariantizing'' terms that all involve commutators, and belong to amplitudes with more legs than the core term. 
For the core term the IBP leads from bulk term to bulk term and is formally identical to the abelian case, so that the
replacement rule holds. But a core term in the effective action appears combined with all its covariantizing terms,
all sharing the same coefficient. The replacement rule induces a change of this coefficient defined through a cycle
$\dot G_{Bi_1i_2} 
\dot G_{Bi_2i_3} 
\cdots
\dot G_{Bi_ni_1}
$
which is multiplied by a $\tr (f_{i_1}f_{i_2}\cdots f_{i_n})$ that
in the effective action corresponds to a $\tr (f^n)$. Consistency is therefore possible only if the
same change of coefficient applies also to all the terms where one or several of the factors $f^{\mu\nu}$
are replaced by a $[A,A]$ term, or where a $\partial$ acting on some $F$ is replaced by a $[A,F]$. 
This settles the spinor loop case. For the gluon loop case, it is sufficient to remember that the corresponding
master formula before taking the limit $C\to\infty$ and the projector sum $\sum_{p=P,A}$, eq. (\ref{gluonmaster}),
is still isomorphic to the spinor loop one (\ref{supermaster}). 

Based on this general validity of the off-shell replacement rules and the general IBP algorithm
developed in \cite{91} we anticipate that with the method presented here a first 
explicit calculation of the four-point
vertex should be well in reach. 

Less straightforward but very interesting would be the extension of the formalism presented here to 
the gravitational case. The one-loop three-graviton vertices have already been
extensively studied for their conformal properties (see \cite{cdms} and references therein). 
As far as external gravitons are concerned,  suitable string-inspired representations 
exist already for the (off-shell) one-loop $N$ - graviton amplitudes with a scalar or spinor loop \cite{baszir,bacozi} as
well as for a photon loop \cite{babegi2005-1,babegi2005-2}.  However,  a suitable IBP procedure still remains to be developed. 

\medskip
\noindent
Acknowledgements: We thank A. Davydychev for a number of helpful discussions, as well as explanations on
\cite{daosta-3gluonD,daossa-3gluonDm} and comments on the manuscript. Discussions and correspondence
with R. Alkofer, F. Bastianelli, C. Corian\`o and J.~M. Cornwall are gratefully acknowledged.
Both authors thank the Physics Department of Bologna University for hospitality, 
and C.~S. also the Albert-Einstein-Institut, Potsdam. 
The work of C.~S. was supported by CONACYT grant CB 2008 101353 
and the work of N.~A. through a CONACYT PhD fellowship.

\begin{appendix}

\section{Summary of Conventions}
\label{conv}

\renewcommand{\theequation}{A.\arabic{equation}}
\setcounter{equation}{0}
\vskip10pt

We work with the $(-+++)$ metric. 
The nonabelian covariant derivative is
$D_\mu\equiv \partial_\mu+ig A^a_\mu  T^a$,
with $[T^a,T^b] = i f^{abc}T^c$. The adjoint
representation is given by $(T^a)^{bc}=-i f^{abc}$.
The normalization of the generators is 
${\rm tr}(T^aT^b) = C(r) \delta^{ab}$, where
for $SU(N)$ one has $C(N)=\frac{1}{2}$ for 
the fundamental and $C(G)= N$ for the adjoint representation. 

\end{appendix}

 \end{document}